\documentclass[aps,prd,superscriptaddress,amsmath,amsfont,amssymb,preprint,nofootinbib]{revtex4}
\pdfoutput=1
\allowdisplaybreaks
\usepackage{graphicx}
\usepackage{ bbold }
\usepackage{subfigure}
\usepackage{slashed}
\usepackage{mathtools}
\usepackage[english]{babel}
\newcommand{\op}{{\cal O}}

\newcommand{\C}{{\cal C}}
\newcommand{\Q}{{\cal Q}}
\newcommand{\mcO}{\mathcal{O}}

\newcommand{\todo}[1]{{\color{red} \ifmmode\else[todo]\fi #1}}
\usepackage[usenames,dvipsnames]{xcolor}
     \definecolor{hgreen}{rgb}{0,.3,0}
     \definecolor{hred}{rgb}{.3,0,0}
     \definecolor{hblue}{rgb}{0,0,.3}
     \definecolor{LightGray}{gray}{0.95}
\usepackage[backref=page,
            colorlinks=true,
            linkcolor=hblue,
            citecolor=hgreen,
            filecolor=hblue,
            urlcolor=hred]{hyperref}

\usepackage{fancyvrb}
\usepackage{upquote}

\renewcommand*{\backref}[1]{}

\newcommand{\mc}[1]{\mathcal{#1}}

\newcommand{\beq}{\begin{equation} }
\newcommand{\eeq}{\end{equation}} 
\newcommand{\bi}{\begin{itemize} }
\newcommand{\ei}{\end{itemize} }

\definecolor{Red}{rgb}{1.,0.,0.}
\definecolor{Grn}{rgb}{0.,0.75,0.}
\definecolor{Blu}{rgb}{0.,0.,1.}
\definecolor{DrkGrn}{HTML}{00AA00}

\DeclareMathOperator{\Tr}{Tr}

\definecolor{pan624}{rgb}{0.482,0.635,0.588} 
\definecolor{pan576}{rgb}{0.412,0.569,0.231} 
\definecolor{pan129}{rgb}{0.961,0.812,0.278}
\definecolor{pan5405}{rgb}{0,0.129,0.278} 
\definecolor{shadecolor}{rgb}{0.482,0.635,0.588}
\definecolor{mygray}{HTML}{666666}
\definecolor{x11steelblue}{HTML}{4682B4}
\definecolor{x11firebrick}{HTML}{B22222}
\definecolor{x11forestgreen}{HTML}{228B22}

\usepackage{tikz}
\tikzstyle{every picture}+=[remember picture]
\usetikzlibrary{shapes.geometric}
\usetikzlibrary{calc}
\usetikzlibrary{decorations.markings}
\usetikzlibrary{decorations.text}
\usetikzlibrary{patterns}
\usetikzlibrary{backgrounds}
\usetikzlibrary{positioning}
\tikzstyle arrowstyle=[scale=2]
\tikzstyle directed=[postaction={decorate,decoration={markings,
		mark=at position 0.6 with {\arrow[arrowstyle]{>}}}}]
\tikzstyle rarrow=[postaction={decorate,decoration={markings,
		mark=at position 0.999 with {\arrow[arrowstyle]{>}}}}]

\newcommand{\lrpartial}{\negthickspace\stackrel{\leftrightarrow}{\partial}\negthickspace{}}
\newcommand{\lpartial}{\negthickspace\stackrel{\leftarrow}{\partial}\negthickspace{}}
\newcommand{\rpartial}{\negthickspace\stackrel{\rightarrow}{\partial}\negthickspace{}}

\newcommand{\slpartial}{\negthickspace\stackrel{\leftarrow}{\slashed{\partial}}\negthickspace{}}
\newcommand{\srpartial}{\negthickspace\stackrel{\rightarrow}{\slashed{\partial}}\negthickspace{}}

\newcommand{\vasq}{\vec{v}_a^{\,\,2}}

\graphicspath{{./figs/}}

\newcommand{\ncdot}{\negthinspace \cdot \negthinspace}

\begin{document}

\title{From quarks to nucleons in dark matter direct detection}

\def\Cincy{Department of Physics, University of Cincinnati, Cincinnati, Ohio 45221,USA}
\def\UCSD{Department of Physics, University of California-San Diego, La Jolla, CA 92093, USA}
\def\Mainz{PRISMA Cluster of Excellence \& Mainz Institute for Theoretical
Physics, Johannes Gutenberg University, 55099 Mainz, Germany}
\def\TUD{Fakult\"at f\"ur Physik, TU Dortmund, D-44221 Dortmund, Germany} 
\def\CERN{CERN, Theory Division, CH-1211 Geneva 23, Switzerland}
\def\Oxford{Rudolf Peierls Centre for Theoretical Physics, University of Oxford OX1 3NP Oxford, United Kingdom}

\author{\textbf{Fady Bishara}}
\email{fady.bishara AT physics.ox.ac.uk}
\affiliation{\Oxford}

\author{\textbf{Joachim Brod}}
\email{joachim.brod AT tu-dortmund.de}
\affiliation{\TUD}

\author{\textbf{Benjamin Grinstein}}
\email{bgrinstein AT ucsd.edu}
\affiliation{\UCSD}

\author{\textbf{Jure Zupan}} 
\email{zupanje AT ucmail.uc.edu}
\affiliation{\Cincy}
\affiliation{\CERN}

\date{\today}

\begin{abstract}
We provide expressions for the nonperturbative matching of the
effective field theory describing dark matter interactions with quarks
and gluons to the effective theory of nonrelativistic dark matter
interacting with nonrelativistic nucleons. We give expressions of
leading and subleading order in chiral counting.  In general, a single
partonic operator matches onto several nonrelativistic operators
already at leading order in chiral counting. Keeping only one operator
at the time in the nonrelativistic effective theory thus does not
properly describe the scattering in direct detection. The matching of
the axial--axial partonic level operator, as well as the matching of
the operators coupling DM to the QCD anomaly term, include naively
momentum suppressed terms. However, these are still of leading chiral
order due to pion poles and can be numerically important.
\end{abstract}

\pacs{--pacs--}

\preprint{DO-TH 17/10}
\preprint{OUTP-17-07P}
\preprint{CERN-TH-2017-157}

\maketitle
\tableofcontents


\section{Introduction}
\label{sec:Intro}
\begin{figure}
\includegraphics[scale=0.8]{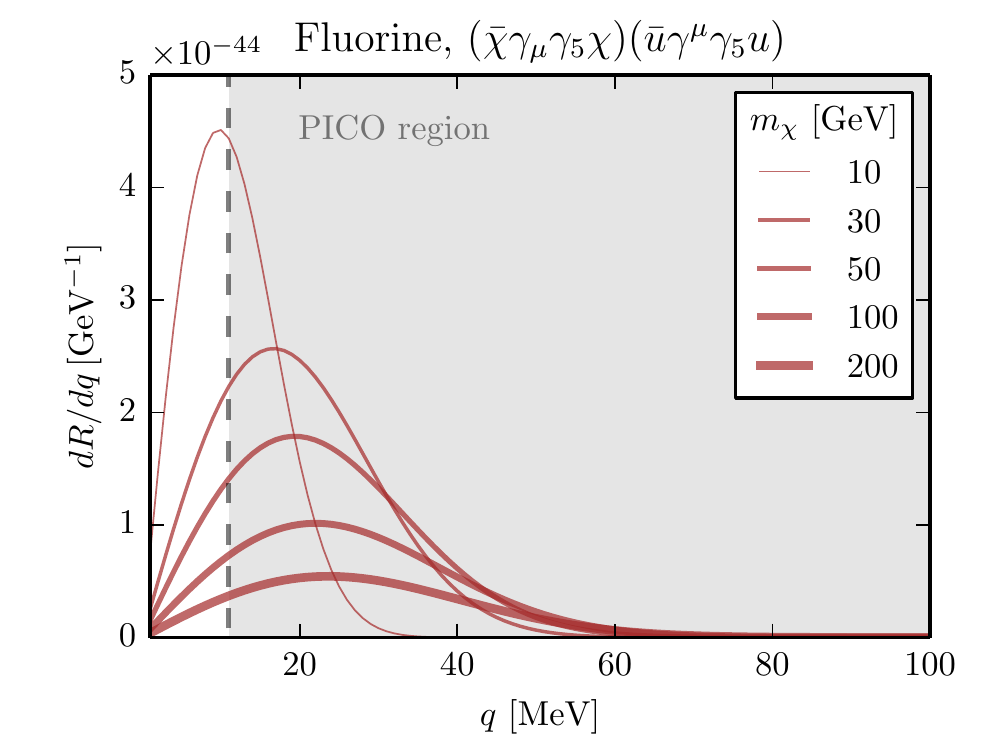}\hspace*{0.2cm}
\includegraphics[scale=0.8]{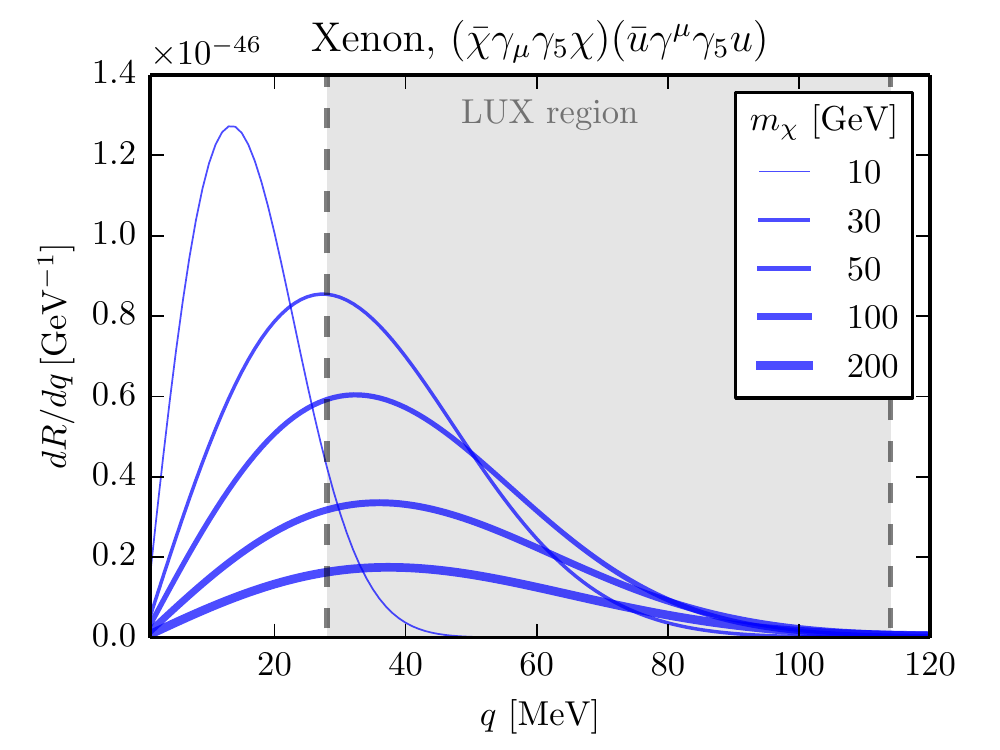}
\caption{The momentum exchange distributions for DM scattering on a
  representative light nucleus, ${}^{19}$F, (left) and heavy nucleus,
  Xe, (right) for spin-dependent scattering. The Wilson coefficient of
  the operator is set to $(1\,\text{TeV})^{-2}$ in both cases and we
  summed the contributions of the xenon isotopes weighted by their
  natural abundances.  The curves of different thicknesses correspond
  to different dark matter masses in GeV as shown in the plot legends.
  The approximate experimental thresholds are denoted by dashed
  vertical lines. For fluorine, we use the PICO threshold region
  $E_R>3.3$ keV~\cite{Amole:2017dex} while for LUX, we use the
  approximate region $E_R\in[3,50]$ keV~\cite{Akerib:2017kat}.  }
\label{fig:dRdq}
\end{figure}

Dark Matter (DM) direct detection, where DM scatters on a target
nucleus, is well described by Effective Field Theory (EFT)
\cite{Bishara:2016hek,Fan:2010gt, Fitzpatrick:2012ix,
  Fitzpatrick:2012ib, Anand:2013yka,DelNobile:2013sia,
  Barello:2014uda, Hill:2014yxa,Catena:2014uqa, Kopp:2009qt,
  Hill:2013hoa, Hill:2011be, Kurylov:2003ra, Pospelov:2000bq,
  Bagnasco:1993st, Cirigliano:2012pq, Hoferichter:2015ipa,
  Hoferichter:2016nvd}, which is essential to compare results of
different direct detection experiments \cite{Catena:2016hoj}. The
maximal momentum exchange between DM and the nucleus is $q_{\rm
  max}\lesssim 200$ MeV, see Fig.~\ref{fig:dRdq}. This means that one
is able to use chiral counting, with an expansion parameter
$q/\Lambda_{\rm ChEFT}\lesssim 0.3$ to organize different
contributions in the nucleon EFT for each of the operators coupling DM
to quarks and gluons \cite{Weinberg:1990rz, Weinberg:1991um,
  Bedaque:2002mn, Epelbaum:2008ga, Epelbaum:2010nr, Cirigliano:2012pq,
  Bishara:2016hek}. In this paper we rewrite the leading-order (LO)
results in the chiral expansion of Ref.~\cite{Bishara:2016hek} in
terms of single-nucleon form factors.  We also extend these results to
higher orders in the $(q/\Lambda_{\rm ChEFT})^2$ expansion up to the
order where two-nucleon currents are expected to become important (for
the discussion of two-nucleon currents and numerical estimates see
\cite{Cirigliano:2012pq, Gazda:2016mrp, Korber:2017ery,
  Hoferichter:2015ipa, Hoferichter:2016nvd}). We give several
numerical examples illustrating that it is not always justified to use
momentum-independent coefficients in the nonrelativistic EFT for DM
interactions with nucleons~\cite{Fitzpatrick:2012ix,
  Fitzpatrick:2012ib, Anand:2013yka}. One needs to include the
light-meson poles when DM couples to axial quark current or to the QCD
anomaly term, to capture the leading effects of the strong
interactions. Similarly, assuming that only one of the norelativistic
EFT operators contributes may be equally hard to justify in a more
complete UV theory. From a particle-physics point of view it is easier
to interpret the results of DM direct detection experiments if one
uses an EFT where DM couples to quark and gluons.

Our starting point is thus the interaction Lagrangian between DM and
the SM quarks, gluons, and photon, given by a sum of higher dimension
operators,
\begin{equation}\label{eq:lightDM:Lnf5}
{\cal L}_\chi=\sum_{a,d}
\hat \C_{a}^{(d)} {\cal Q}_a^{(d)}, 
\qquad {\rm where}\quad 
\hat \C_{a}^{(d)}=\frac{\C_{a}^{(d)}}{\Lambda^{d-4}}\,.
\end{equation}
Here the $\C_{a}^{(d)}$ are dimensionless Wilson coefficients, while
$\Lambda$ can be identified with the mass of the mediators between DM
and the SM (for couplings of order unity). The sums run over the
dimensions of the operators, $d=5,6,7$ and the operator labels,
$a$. Depending on the operator, the label `$a$' either denotes an
operator number or a number and a flavor index if the operator
contains a SM fermion bilinear. We keep all the operators of
dimensions five and six, and all the operators of dimension seven that
couple DM to gluons. Among the dimension-seven operators that couple
DM to quarks we exclude from the discussion the operators that are
additionally suppressed by derivatives but have otherwise the same
chiral structure as the dimension-six operators (for the treatment of
these operators see \cite{Brod:2017bsw}).

There are two dimension-five operators,
\begin{equation}
\label{eq:dim5:nf5:Q1Q2:light}
{\cal Q}_{1}^{(5)} = \frac{e}{8 \pi^2} (\bar \chi \sigma^{\mu\nu}\chi)
 F_{\mu\nu} \,, \qquad {\cal Q}_2^{(5)} = \frac{e }{8 \pi^2} (\bar
\chi \sigma^{\mu\nu} i\gamma_5 \chi) F_{\mu\nu} \,,
\end{equation}
where $F_{\mu\nu}$ is the electromagnetic field strength tensor and
$\chi$ is the DM field, assumed here to be a Dirac particle. The
magnetic dipole operator $\Q_1^{(5)}$ is CP even, while the electric
dipole operator $\Q_2^{(5)}$ is CP odd. 
The dimension-six operators are
\begin{align}
{\cal Q}_{1,q}^{(6)} & = (\bar \chi \gamma_\mu \chi) (\bar q \gamma^\mu q)\,,
 &{\cal Q}_{2,q}^{(6)} &= (\bar \chi\gamma_\mu\gamma_5 \chi)(\bar q \gamma^\mu q)\,, \label{eq:dim6EW:Q1Q2:light}
  \\ 
{\cal Q}_{3,q}^{(6)} & = (\bar \chi \gamma_\mu \chi)(\bar q \gamma^\mu \gamma_5 q)\,,
  & {\cal Q}_{4,q}^{(6)}& = (\bar
\chi\gamma_\mu\gamma_5 \chi)(\bar q \gamma^\mu \gamma_5 q)\,,\label{eq:dim6EW:Q3Q4:light}
\end{align}
and we also include a subset of the dimension-seven operators,
namely\footnote{Note that the definition of the operator ${\cal
    Q}_{8,q}^{(7)}$ differs by a sign from the definition used
  in~\cite{Bishara:2016hek}.} 
\begin{align}
{\cal Q}_1^{(7)} & = \frac{\alpha_s}{12\pi} (\bar \chi \chi)
 G^{a\mu\nu}G_{\mu\nu}^a\,, 
 & {\cal Q}_2^{(7)} &= \frac{\alpha_s}{12\pi} (\bar \chi i\gamma_5 \chi) G^{a\mu\nu}G_{\mu\nu}^a\,,\label{eq:dim7:Q1Q2:light}
 \\
{\cal Q}_3^{(7)} & = \frac{\alpha_s}{8\pi} (\bar \chi \chi) G^{a\mu\nu}\widetilde
 G_{\mu\nu}^a\,, 
& {\cal Q}_4^{(7)}& = \frac{\alpha_s}{8\pi}
(\bar \chi i \gamma_5 \chi) G^{a\mu\nu}\widetilde G_{\mu\nu}^a \,, \label{eq:dim7:Q3Q4:light}
\\
{\cal Q}_{5,q}^{(7)} & = m_q (\bar \chi \chi)( \bar q q)\,, 
&{\cal
  Q}_{6,q}^{(7)} &= m_q (\bar \chi i \gamma_5 \chi)( \bar q q)\,,\label{eq:dim7EW:Q5Q6:light}
  \\
{\cal Q}_{7,q}^{(7)} & = m_q (\bar \chi \chi) (\bar q i \gamma_5 q)\,, 
&{\cal Q}_{8,q}^{(7)} & = m_q (\bar \chi i \gamma_5 \chi)(\bar q i \gamma_5
q)\,, \label{eq:dim7EW:Q7Q8:light}  
 \\
{\cal Q}_{9,q}^{(7)} & = m_q (\bar \chi \sigma^{\mu\nu} \chi) (\bar q \sigma_{\mu\nu} q)\,, 
&{\cal Q}_{10,q}^{(7)} & = m_q (\bar \chi  i \sigma^{\mu\nu} \gamma_5 \chi)(\bar q \sigma_{\mu\nu}
q)\,. \label{eq:dim7EW:Q9Q10:light} 
\end{align}
Here $G_{\mu\nu}^a$ is the QCD field strength tensor, while
$\widetilde G_{\mu\nu} = \frac{1}{2}\varepsilon_{\mu\nu\rho\sigma}
G^{\rho\sigma}$ is its dual, and $a=1,\dots,8$ are the adjoint color
indices. Moreover, $q=u,d,s$ denote the light quarks (we limit
ourselves to flavor conserving operators). Note that we include two
more dimension-seven operators than in~\cite{Bishara:2016hek}, so that
we have all the operators included in~\cite{Goodman:2010ku}. The
remaining dimension-seven operators coupling DM to quarks are listed
in \cite{Brod:2017bsw}, while the effect of dimension-seven operators
coupling DM to photons is discussed in \cite{Ovanesyan:2014fha}.
There are also the leptonic equivalents of the operators ${\cal
  Q}_{1,q}^{(6)},\ldots, {\cal Q}_{4,q}^{(6)}$, and ${\cal
  Q}_{5,q}^{(7)},\ldots, {\cal Q}_{10,q}^{(7)}$, with $q\to \ell$.

\begin{figure}
\includegraphics[scale=0.5]{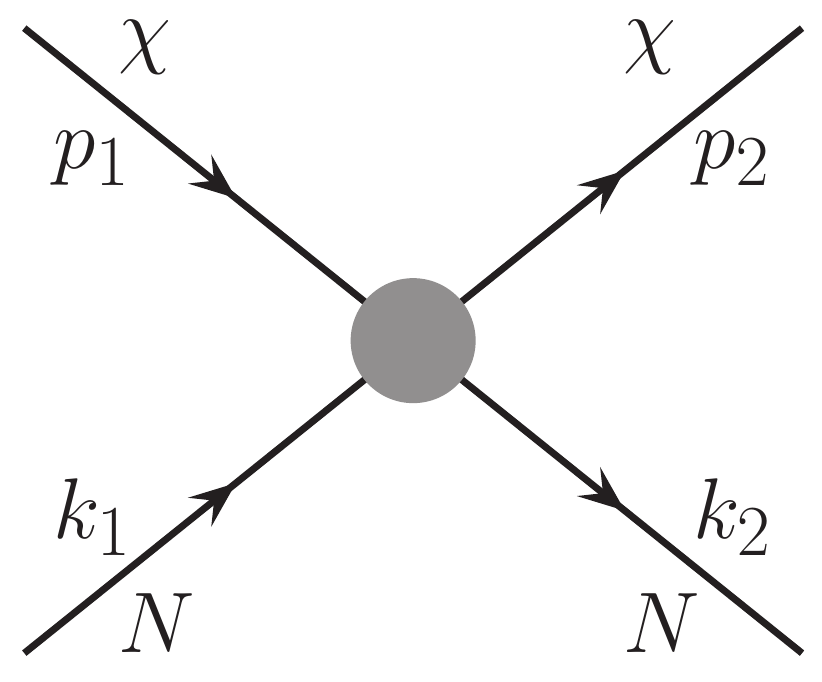}~~~~~~~~~~
\caption{The kinematics of DM scattering on nucleons,
  $\chi(p_1)N(k_1)\to \chi(p_2) N(k_2)$.  }
         \label{fig:scattering_kin}
\end{figure}

The aim of this paper is to provide compact expressions for the
non-perturbative matching at $\mu\simeq 2$ GeV between the EFT with
three quark flavors, given by Eq.~\eqref{eq:lightDM:Lnf5}, and the
theory of DM interacting with nonrelativistic nucleons, given by
\begin{equation}\label{eq:LNR}
{\cal L}_{\rm NR}=\sum_{i,N} c_i^N(q^2) \op_i^N.
\end{equation}
For each operator the matching is done using the heavy baryon chiral
perturbation theory expansion~\cite{Jenkins:1990jv} up to the order
for which the scattering amplitudes are still parametrically dominated
by single-nucleon currents. The relevant Galilean-invariant operators
with at most two derivatives are
\begin{align}
\label{eq:O1pO2p}
{\mathcal O}_1^N&= \mathbb{1}_\chi \mathbb{1}_N\,,
&{\mathcal O}_2^N&= \big(v_\perp\big)^2 \, \mathbb{1}_\chi \mathbb{1}_N\,,
\\
\label{eq:O3pO4p}
{\mathcal O}_3^N&= \mathbb{1}_\chi \, \vec S_N\cdot \Big(\vec v_\perp\negthickspace \times \frac{i\vec q}{m_N}\Big) \,,
&{\mathcal O}_4^N&= \vec S_\chi \cdot \vec S_N \,,
\\
\label{eq:O5pO6p}
{\mathcal O}_5^N&= \vec S_\chi \cdot \Big(\vec v_\perp \times \frac{i\vec q}{m_N} \Big) \, \mathbb{1}_N \,,
&{\mathcal O}_6^N&= \Big(\vec S_\chi \cdot \frac{\vec q}{m_N}\Big) \, \Big(\vec S_N \cdot \frac{\vec q}{m_N}\Big),
\\
\label{eq:O7pO8p}
{\mathcal O}_7^N&= \mathbb{1}_\chi \, \big( \vec S_N \cdot \vec v_\perp \big)\,,
&{\mathcal O}_8^N&= \big( \vec S_\chi \cdot \vec v_\perp \big) \, \mathbb{1}_N\,,
\\
\label{eq:O9pO10p}
{\mathcal O}_9^N&= \vec S_\chi \cdot \Big(\frac{i\vec q}{m_N} \times \vec S_N \Big)\,,
&{\mathcal O}_{10}^N&= - \mathbb{1}_\chi \, \Big(\vec S_N \cdot \frac{i\vec q}{m_N} \Big)\,,
\\
\label{eq:O11pO12p}
{\mathcal O}_{11}^N&= - \Big(\vec S_\chi \cdot \frac{i\vec q}{m_N} \Big) \, \mathbb{1}_N \,,
&{\mathcal O}_{12}^N&= \vec S_\chi \cdot \Big( \vec S_N \times \vec v_\perp \Big) \,,
\\
\label{eq:O13pO14p}
{\mathcal O}_{13}^N&= -\Big(\vec S_\chi \cdot \vec v_\perp \Big) \, \Big(\vec S_N\cdot \frac{i\vec q}{m_N} \Big) \,,
&{\mathcal O}_{14}^N&= -\Big(\vec S_\chi \cdot \frac{i\vec q}{m_N}  \Big) \, \Big(\vec S_N\cdot \vec v_\perp  \Big) \,,
\end{align}
and in addition
\begin{equation}
\op_{2b}^N=\big(\vec S_N\cdot \vec v_\perp\big)\big(\vec S_\chi\cdot
\vec v_\perp\big)\,,
\end{equation}
where $N=p,n$. At next-to-leading order (NLO) we also need one
operator with three derivatives,
\begin{equation}
\label{eq:O15p}
{\mathcal O}_{15}^N= -\Big(\vec S_\chi \cdot \frac{\vec q}{m_N} \Big) \, \Big(\big(\vec S_N\times \vec v_\perp \big)\cdot \frac{\vec q}{m_N} \Big)\,. 
\end{equation}
Our definition of momentum exchange differs from~\cite{Anand:2013yka} by a minus sign, cf. Fig. \ref{fig:scattering_kin},
\begin{equation}
\vec q = \vec k_2-\vec k_1=\vec p_1 -\vec p_2\,, \qquad \vec v_\perp=\big(\vec p_1+\vec p_2\big)/(2{m_\chi})-\big(\vec k_1+\vec k_2\big)/(2{m_N})\,,
\end{equation}
while the operators coincide with those defined
in~\cite{Anand:2013yka}. 
Each insertion of $\vec q$ is accompanied by a factor of $1/m_N$, so
that all of the above operators have the same dimensionality.

This paper is organized as follows: in
Section~\ref{subsec:nonpert:chiral} we give the matching conditions
for fermionic DM and in Section~\ref{sec:scalarDM} for scalar DM,
while in Section~\ref{sec:examples} we present several examples
illustrating the importance of keeping all terms of the same chiral
order. Section \ref{sec:conclusions} contains our conclusions. The
numerical values of the form factors are collected in
Appendix~\ref{App:form:factors}, Appendix~\ref{app:NR} contains the
nonrelativistic expansion of the fermionic DM and nucleon currents,
Appendix \ref{sec:NLOva} the extended NLO operator basis,
Appendix~\ref{app:NR:scalar} the NLO results for scalar DM, while
Appendix~\ref{sec:anand} gives the results for fermionic DM in terms
of coefficients of the nonrelativistic operators.

\section{Fermionic dark matter}
\label{subsec:nonpert:chiral}

\begin{figure}
\includegraphics[scale=1]{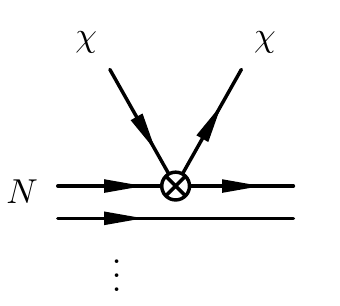}\hspace*{1cm}
\includegraphics[scale=1]{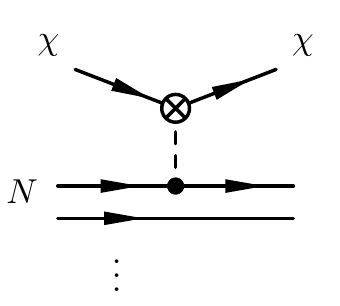}\hspace*{1cm}
\includegraphics[scale=1]{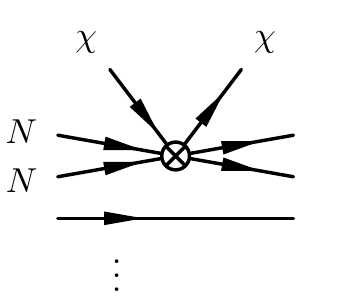}
\caption{The chirally leading diagrams for DM-nucleus scattering (the
  first and second diagrams), and a representative diagram for
  two-nucleon scattering (the third diagram). The effective
  DM--nucleon and DM--meson interactions are denoted by a circle, the
  dashed line denotes a pion, and the dots represent the remaining
  $A-2$ nucleon lines.  }
         \label{fig:LOChPT}
\end{figure}

The hadronization of operators $\Q_{1,q}^{(6)},\dots,\Q_{10,q}^{(7)}$,
in Eqs.~\eqref{eq:dim6EW:Q1Q2:light}-\eqref{eq:dim7EW:Q9Q10:light}
leads at LO in the chiral expansion only to single-nucleon
currents~\cite{Bishara:2016hek}. The scattering of DM on a nucleus
with mass number $A$ is given by a sum of $A$-nucleon irreducible
amplitudes with one DM current insertion. These amplitudes scale as
$\mathcal{M}_{A,\chi}\sim(q/\Lambda_\textrm{ChEFT})^\nu$ where the
power counting exponent $\nu$ is given explicitly
in~\cite{Bishara:2016hek}. This counting was first derived by Weinberg
in~\cite{Weinberg:1991um} -- see also~\cite{Bedaque:2002mn,
  Cirigliano:2012pq}.  In the case of our EFT basis, the matrix
elements of the operators scale as $q^{\nu_{\rm LO}}$, with
\cite{Bishara:2016hek, Brod:2017bsw}
\begin{equation}
\label{eq:chiral:scaling}
\begin{matrix*}[l]
[\Q_{1,q}^{(6)}]_{\rm LO}\sim 1, & \quad [\Q_{2,q}^{(6)}]_{\rm LO}\sim q, & \quad [\Q_{3,q}^{(6)}]_{\rm LO}\sim q, & \quad [\Q_{4,q}^{(6)}]_{\rm LO}\sim 1,
\\
[\Q_{1}^{(7)}]_{\rm LO}\sim 1,  & \quad[\Q_{2}^{(7)}]_{\rm LO}\sim q, &\quad [\Q_{3}^{(7)}]_{\rm LO}\sim q, & \quad [\Q_{4}^{(7)}]_{\rm LO}\sim q^2,
\\
[\Q_{5,q}^{(7)}]_{\rm LO}\sim q^2, &\quad [\Q_{6,q}^{(7)}]_{\rm LO}\sim q^3, &\quad [\Q_{7,q}^{(7)}]_{\rm LO}\sim q, & \quad [\Q_{8,q}^{(7)}]_{\rm LO}\sim q^2, 
\\
[\Q_{9,q}^{(7)}]_{\rm LO}\sim 1, & \quad [\Q_{10,q}^{(7)}]_{\rm LO}\sim q, & \quad &\quad
\end{matrix*}
\end{equation}
counting $m_q\sim m_\pi^2\sim q^2$, and not displaying a common
scaling factor.  The LO contributions are either due to scattering of
DM on a single nucleon (the first diagram in Fig.~\ref{fig:LOChPT}),
or on a pion that attaches to the nucleon (the second diagram), or
both.
The contributions from DM scattering on two-nucleon currents arise at
${\mathcal O}(q^{\nu_{\rm LO}+1})$ for ${\cal O}_{2,q}^{(6)}$, ${\cal
  O}_{5,q}^{(7)}$, and ${\cal O}_{6,q}^{(7)}$, at ${\mathcal
  O}(q^{\nu_{\rm LO}+2})$ for ${\cal O}_{1,q}^{(6)} $, and at
${\mathcal O}(q^{\nu_{\rm LO}+3})$ for all the other operators.
Up to these orders, the hadronization of the operators
$\Q_{1,q}^{(6)},\dots,\Q_{10,q}^{(7)}$ can thus be described by using
form factors for single-nucleon currents.

The form factors are given by
\begin{align}
\label{vec:form:factor}
\langle N'|\bar q \gamma^\mu q|N\rangle&=\bar u_N'\Big[F_1^{q/N}(q^2)\gamma^\mu+\frac{i}{2m_N}F_2^{q/N}(q^2) \sigma^{\mu\nu}q_\nu\Big]u_N\,,
\\
\label{axial:form:factor}
\langle N'|\bar q \gamma^\mu \gamma_5 q|N\rangle&=\bar u_N'\Big[F_A^{q/N}(q^2)\gamma^\mu\gamma_5+\frac{1}{2m_N}F_{P'}^{q/N}(q^2) \gamma_5 q^\mu\Big]u_N\,,
\\
\label{scalar:form:factor}
\langle N'| m_q \bar q   q|N\rangle&= F_S^{q/N} (q^2)\, \bar u_N' u_N\,,
\\
\label{pseudoscalar:form:factor}
\langle N'| m_q \bar q  i \gamma_5 q|N\rangle&= F_P^{q/N} (q^2)\, \bar u_N' i \gamma_5 u_N\,,
\\
\label{CPeven:gluonic:form:factor}
\langle N'| \frac{\alpha_s}{12\pi} G^{a\mu\nu}G^a_{\mu\nu} |N\rangle&= F_G^{N} (q^2)\, \bar u_N' u_N\,,
\\
\label{CPodd:gluonic:form:factor}
\langle N'| \frac{\alpha_s}{8\pi} G^{a\mu\nu}\tilde G^a_{\mu\nu}|N\rangle&= F_{\tilde G}^{N} (q^2)\, \bar u_N' i \gamma_5 u_N\,,
\\
\label{tensor:form:factor}
\begin{split}
\langle N'|m_q \bar q \sigma^{\mu\nu} q |N\rangle&=  \bar u_N'\Big[F_{T,0}^{q/N} (q^2)\,  \sigma^{\mu\nu} +\frac{i}{2 m_N} \gamma^{[\mu}q^{\nu]} F_{T,1}^{q/N} (q^2) 
\\
&\qquad \qquad+ \frac{i}{m_N^2} q^{[\mu}k_{12}^{\nu]} F_{T,2}^{q/N} (q^2) \Big] u_N\,,
\end{split}
\end{align}
where we have suppressed the dependence of nucleon states on their
momenta, i.e. $\langle N'|\equiv\langle N(k_2)| $, $| N\rangle\equiv |
N(k_1)\rangle $, and similarly, $\bar u_N'\equiv \bar u_N(k_2)$, $u_N\equiv
u_N(k_1)$. The momentum exchange is $q^\mu=k_2^\mu-k_1^\mu$, while
$k_{12}^\mu=k_1^\mu+k_2^\mu$.  The form factors $F_i$ are functions of
$q^2$ only.

The axial current, the pseudoscalar current, and the CP-odd gluonic
current receive contributions from light pseudoscalar meson exchanges
corresponding to the second diagram in Fig.~\ref{fig:LOChPT}. For
small momenta exchanges, $q\sim m_\pi$, one can expand the form
factors in $q^2$, as 
\begin{align}
\label{eq:F_PP'}
F_{i}^{q/N}(q^2)&=\overbrace{\frac{m_N^2}{m_\pi^2-q^2} a_{i,\pi}^{q/N}+\frac{m_N^2}{m_\eta^2-q^2} a_{i,\eta}^{q/N}}^{\rm LO}+\overbrace{b_{i}^{q/N}}^{\rm NLO}
+\cdots, \quad i=P,P',
\\
\label{eq:F_tildeG}
F_{\tilde G}^{N}(q^2)&=
\underbrace{\frac{q^2}{m_\pi^2-q^2} a_{\tilde G,\pi}^{N}+\frac{q^2}{m_\eta^2-q^2} a_{\tilde G,\eta}^{N}+b_{\tilde G}^{N}
}_{\rm LO}+\underbrace{c_{\tilde G}^{N} q^2}_{\rm NLO}+\cdots,
\end{align}
where we kept both the pion and eta poles and denoted the order of the
various terms in chiral counting. The coefficients $a_i,b_i,c_i$ are
momentum-independent constants. Note that the pion and eta poles for
the $G\tilde G$ operator are suppressed by $q^2$ and are thus of the
same chiral order as the constant term, $b_{\tilde G}^{N}$. All the
other form factors do not have a light pseudoscalar pole and can be
Taylor expanded\footnote{We assume that the NLO terms involving chiral
  logarithms of the form $(m_\pi^2-q^2)\log(m_\pi^2-q^2)$ were also
  expanded in $q^2$. This may give an effective expansion parameter
  $q^2/(\Lambda_{\rm EFT})^2$ with $\Lambda_{\rm EFT}$ between $m_\pi$
  and $4\pi f$; however, numerically it is found to be closer to the
  latter, see Appendix~\ref{App:form:factors}.} around $q^2=0$,
\begin{equation}
\label{eq:Fi}
F_{i}^{q/N}(q^2)=\underbrace{F_{i}^{q/N}(0)}_{\rm
  LO}+\underbrace{F_{i}^{\prime \, q/N}(0)q^2}_{\rm NLO}+\cdots\,,
\end{equation}
where the prime on $F$ denotes a derivative with respect to $q^2$.
The values of $F_i^{q/N}(0)$, $F_i^{\prime \, q/N}(0)$, and $a_i, b_i,
c_i$ are collected in Appendix~\ref{App:form:factors}.

The size of the form factors that do not have light-meson poles are,
at zero recoil,
\begin{align}
&F_{1,2}^{q/N}(0)\,, F_{A}^{q/N}(0) \sim {\mathcal O(1)}\,,&&  F_{1,2}^{s/N}(0)\,, F_{A}^{s/N}(0) \sim {\mathcal O}(0-0.05)\,,
\\
&F_S^{q/N}(0)\sim {\mathcal O}(0.03) m_N\,, && F_S^{s/N}(0)\sim {\mathcal O}(0.05) m_N\,,
\\
&F_G^{N}(0)\sim {\mathcal O}(0.1)m_N\,,
\\
& F_{T,0;T,1;T,2}^{q/N}(0)\sim {\mathcal O}(1) m_q, && F_{T,0;T,1;T,2}^{s/N}(0)\lesssim {\mathcal O}(0.001-0.2)  m_s\,.
\end{align}
(only here and in the remainder of the subsection we use the
abbreviation $q=u,d$). The $s$-quark form factors are much smaller,
with the exception of the scalar form factor. Their derivatives at
zero recoil, which enter the NLO expressions, have a typical size
$F_i'(0)/F_i(0)\sim {\mathcal O}(1/m_N^2)$, so that the corresponding
corrections are expected at the level of several percent.

The coefficients of the terms in the form factors that contain the
pion and eta poles, Eqs. \eqref{eq:F_PP'}, \eqref{eq:F_tildeG}, are
approximately of the size
\begin{align}
&a_{P',\pi}^{q/N}\,, a_{P',\eta}^{q/N}\sim {\mathcal O}(1)\,, &&a_{P',\pi}^{s/N}=0\,, && a_{P',\eta}^{s/N}\sim {\mathcal O}(1)\,,
\\
&a_{P,\pi}^{q/N}\,, a_{P,\eta}^{q/N}\sim {\mathcal O}(1) m_q\,, &&a_{P,\pi}^{s/N}=0\,, && a_{P,\eta}^{s/N}\sim {\mathcal O}(1) m_s\,,
\\
& a_{\tilde G,\pi}^N\,, a_{\tilde G,\eta}^N\,, b_{\tilde G}^N\sim {\mathcal O}(1) m_N\,.
\end{align}

\subsection{Leading-order expressions}
We first give the expressions for the nonrelativistic EFT
Lagrangian~\eqref{eq:LNR} at LO in chiral counting. In this case we
only need the values of $a_{i}^{\pi}$, $a_{i}^{\eta}$, $b_{\tilde
  G}^{N}$, and $F_i(0)$.  In addition to taking the hadronic matrix
elements of the quark and gluon currents we also take the
nonrelativistic limit of both the DM currents and the nucleon
currents. The expressions for this last step are collected in
Appendix~\ref{app:NR}.  The chirally leading hadronization of the
dimension-five operators is thus given by
\begin{align}
\begin{split}
\label{eq:Q15}
\Q_{1}^{(5)}\to& -\frac{\alpha}{2\pi} F_1^{N}\Big(
\frac{1}{m_\chi}\op_1^N-4 \frac{m_N}{\vec q\,^2}
\op_5^N\Big)-\frac{2\alpha}{\pi
}\frac{\mu_N}{m_N}\Big(\op_4^N-\frac{m_N^2}{\vec q\,^2}
\op_6^N\Big)+{\mathcal O}(q^2)\,, 
\end{split}
\\
\begin{split}
\label{eq:Q25}
\Q_{2}^{(5)}\to& \frac{2\alpha}{\pi}\frac{m_N}{\vec q\,^2} F_1^N \op_{11}^N+{\mathcal O}(q^2)\,,
\end{split}
\end{align}
with $F_1^N(0)=\delta_{pN}$ the nucleon charge, and $\mu_N$ the
nucleon magnetic moment (see also Appendix
\ref{app:sec:vector:current}). The dimension-six operators hadronize
as 
\begin{align}
\begin{split}
\Q_{1,q}^{(6)}\to& F_1^{q/N}\op_1^N+{\mathcal O}(q^2)\,,
\end{split}
\\
\begin{split}
\Q_{2,q}^{(6)}\to& 2 F_1^{q/N}\op_8^N+ 2\big(F_1^{q/N}+F_2^{q/N}\big) \op_9^N+{\mathcal O}(q^2)\,,
\end{split}
\\
\begin{split}
\label{eq:LO:Q3q6}
\Q_{3,q}^{(6)}\to& -2 F_A^{q/N} \Big(\op_7^N- \frac{m_N}{m_\chi} \op_9^N\Big)+{\mathcal O}(q^2)\,,
\end{split}
\\
\begin{split}
\Q_{4,q}^{(6)}\to& -4 F_A^{q/N} \op_4^N+F_{P'}^{q/N}\op_6^N+{\mathcal O}(q^2)\,,
\end{split}
\end{align}
while the hadronization of the gluonic dimension-seven operators is given by
\begin{align}
\begin{split}\label{eq:Q17:had:LO}
\Q_{1}^{(7)}\to& F_G^{N}\op_1^N+{\mathcal O}(q^2)\,,
\end{split}
\\
\begin{split}\label{eq:Q27:had:LO}
\Q_{2}^{(7)}\to& -\frac{m_N}{m_\chi}F_G^{N}\op_{11}^N +{\mathcal O}(q^3)\,,
\end{split}
\\
\begin{split}\label{eq:Q37:had:LO}
\Q_{3}^{(7)}\to& F_{\tilde G}^{N}\op_{10}^N+{\mathcal O}(q^3)\,,
\end{split}
\\
\begin{split}\label{eq:Q47:had:LO}
\Q_{4}^{(7)}\to& \frac{m_N}{m_\chi}F_{\tilde G}^{N}\op_{6}^N  +{\mathcal O}(q^4)\,.
\end{split}
\end{align}
The hadronization of the dimension-seven operators with quark scalar
currents results in
\begin{align}
\begin{split}\label{eq:Q5q7:had:LO}
\Q_{5,q}^{(7)}\to& F_S^{q/N}\op_1^N+{\mathcal O}(q)\,,
\end{split}
\\
\begin{split}\label{eq:Q6q7:had:LO}
\Q_{6,q}^{(7)}\to& -\frac{m_N}{m_\chi}F_S^{q/N}\op_{11}^N +{\mathcal O}(q^2)\,,
\end{split}
\\
\begin{split}\label{eq:Q7q7:had:LO}
\Q_{7,q}^{(7)}\to& F_{P}^{q/N}\op_{10}^N  +{\mathcal O}(q^3)\,,
\end{split}
\\
\begin{split}\label{eq:Q8q7:had:LO}
\Q_{8,q}^{(7)}\to& \frac{m_N}{m_\chi}F_{P}^{q/N}\op_{6}^N   +{\mathcal O}(q^4)\,,
\end{split}
\end{align}
and for the tensor operators 
\begin{align}
\begin{split}\label{eq:Q9q7:had:LO}
\Q_{9,q}^{(7)}\to&  8 F_{T,0}^{q/N}\op_4^N+ {\mathcal O}(q^2)\,,
\end{split}
\\
\begin{split}\label{eq:Q10q7:had:LO}
\Q_{10,q}^{(7)}\to&  -2 \frac{m_N}{m_\chi} F_{T,0}^{q/N}\op_{10}^N
+ 2 \big(F_{T,0}^{q/N}-F_{T,1}^{q/N}\big)\op_{11}^N
-8 F_{T,0}^{q/N}\op_{12}^N+ {\mathcal O}(q^3)\,.
\end{split}
\end{align}
The nonrelativistic operators have been defined
in~\eqref{eq:O1pO2p}-\eqref{eq:O13pO14p}. In the above expressions all
the form factors $F_i^{q/N}$ are evaluated at $q^2=0$, apart from
$F_{P,P'}^{q/N}$ and $F_{\tilde G}^{N}$, where one needs to keep the
two meson-pole terms in \eqref{eq:F_PP'} and the first three terms in
\eqref{eq:F_tildeG}. The corresponding values of coefficients $c_i^N$
in the nonrelativistic Larangian, Eq.~\eqref{eq:LNR}, are given in
Appendix~\ref{sec:anand}.

Several comments are in order. First of all, in several cases a single
operator describing the DM interactions with quarks and gluons matches
onto more than one nonrelativistic operator in
Eqs.~\eqref{eq:O1pO2p}-\eqref{eq:O15p} already at leading chiral
order. This occurs for 
\begin{align}
\begin{split}
\Q_{1}^{(5)}&=\frac{e}{8 \pi^2} (\bar \chi \sigma^{\mu\nu}\chi)
 F_{\mu\nu} \sim Q_p\mathbb{1}_\chi \mathbb{1}_N/m_\chi +Q_p  \vec S_\chi\cdot (\vec v_\perp \times i \vec q)\mathbb{1}_N /\vec q^{\,\,2}
 \\ &
 \qquad\qquad\qquad\qquad+\mu_N\vec S_\chi \cdot \vec S_N /m_N +\mu_N(\vec S_\chi \cdot \vec q)(\vec S_N\ncdot \vec q)/(m_N \vec q^{\,\,2})\,,
 \end{split}
 \\
\begin{split}
\Q_{2,q}^{(6)}&= (\bar \chi\gamma_\mu\gamma_5 \chi)(\bar q \gamma^\mu q)\, \sim\, \big( \vec S_\chi \cdot \vec v_\perp \big) \, \mathbb{1}_N+ F_{1,2}^{q/N}(0) \vec S_\chi \cdot \big(i\vec q \times \vec S_N \big)/m_N\,,
\end{split}
\\
\begin{split}
\Q_{3,q}^{(6)}&=(\bar \chi \gamma_\mu \chi)(\bar q \gamma^\mu \gamma_5 q)\, \sim\, \Delta q_N \Big[\mathbb{1}_\chi \, \big( \vec S_N \cdot \vec v_\perp \big)- \vec S_\chi \cdot \big(i\vec q \times \vec S_N \big)/m_\chi \Big]\,,
\end{split}
\\
\begin{split}
\label{eq:Q46:schematic}
\Q_{4,q}^{(6)}&=(\bar
\chi\gamma_\mu\gamma_5 \chi)(\bar q \gamma^\mu \gamma_5 q)\, \sim\, \Delta q_N  \vec S_\chi \cdot \vec S_N + \frac{\Delta q_N}{m_\pi^2+\vec q^{\,\,2}}\big(\vec S_\chi \cdot \vec q\,\big) \, \big(\vec S_N \cdot \vec q \,\big)\,,
\end{split}
\\
\begin{split}
\Q_{10,q}^{(7)}&=m_q (\bar \chi  i \sigma^{\mu\nu} \gamma_5 \chi)(\bar q \sigma_{\mu\nu}
q)\, \sim\,  \frac{m_q}{m_\chi} g_{T}^q \mathbb{1}_\chi \, \big(\vec S_N \cdot i\vec q \,\big)
+ \frac{m_q}{m_N}\{  g_{T}^q,F_{T,1}^{q/N}\big\} \big(\vec S_\chi \cdot i\vec q \,\big) \, \mathbb{1}_N
\\
&\qquad\qquad\qquad\qquad\qquad\qquad +m_q g_{T}^q \vec S_\chi \cdot \big( \vec S_N \times \vec v_\perp \big)\,,
\end{split}
\end{align}
where we only show the approximate dependence on the nonperturbative
coefficients (here $Q_p=1$ is the proton charge, while the values of
the axial charge $\Delta q_N$, the form factors $F_{1,2}^{q/N}(0)$ and
the tensor charges, $g_{T}^q$, $F_{T,1}^{q/N}(0)$) are given in
Appendix~\ref{App:form:factors}.

The above results mean that it is not consistent within EFT to perform
the direct detection analysis in the nonrelativistic basis and only
turn on one of the operators $\op_7^N,\op_8^N, \op_9^N$ or
$\op_{12}^N$, as they always come accompanied with other
nonrelativistic operators, regardless of the UV operator that couples
DM to quarks and gluons. On the other hand, the spin-independent
operator $\op_1^N$ as well as the spin-dependent operator $\op_4^N$
can arise by themselves from $\Q_{1,q}^{(6)}, \Q_1^{(7)},
\Q_{5,q}^{q/N}$ and from $\Q_{9,q}^{(7)}$, respectively. Similarly,
$\op_6^N$, $\op_{10}^N$, and $\op_{11}^N$ arise as the only leading
operators in the nonrelativistic reduction of $\Q_{8,q}^{(7)}$,
$\Q_{3}^{(7)}$ or $\Q_{7,q}^{(7)}$, and $\Q_{2}^{(7)}$ or
$\Q_{6,q}^{(7)}$, respectively.
 
While it is true that the spin-dependent operator $\op_4^N$ can arise
from the tensor-tensor operator $\Q_{9,q}^{(7)}$, this contribution
would be of two-loop order in a perturbative UV theory of DM. The
axial-axial operator $Q_{4,q}^{(6)}$, on the other hand, also leads to
spin-dependent scattering and will arise at tree level. Therefore it
will, if generated, typically dominate over $\Q_{9,q}^{(7)}$.  The
induced spin-dependent scattering arises from both the $\op_4^N = \vec
S_\chi \cdot \vec S_N$ and $\op_6^N = \big(\vec S_\chi \cdot \vec q
\,\big) \, \big(\vec S_N \cdot \vec q \,\big)$ operators. While the
latter is ${\mathcal O}(q^2)$ suppressed, it is simultaneously
enhanced by $1/(m_\pi^2+\vec q^{\,\,2})$ so that in general the two
contributions are of similar size (for scattering on heavy nuclei). In
this case, again, one cannot perform the direct detection analysis
with just $\op_4^N$ or just $\op_6^N$. The same is true for the
operators $\Q_{2,q}^{(6)}$, $\Q_{3,q}^{(6)}$, and $\Q_{10,q}^{(7)}$
that each match at leading order in chiral counting to at least two
nonrelativistic operators. Therefore, a correct LO description of the
DM scattering rate cannot be achieved by using only one
nonrelativistic operator at a time.  We explore this quantitatively in
Section \ref{sec:examples}, also distinguishing the cases of light and
heavy nuclei.

\subsection{Subleading corrections}
\label{sec:subleading}
We discuss next the NLO corrections to the nonrelativistic reduction
of the
operators~\eqref{eq:dim6EW:Q1Q2:light}-\eqref{eq:dim7EW:Q9Q10:light}. The
explicit expressions are given in Appendix~\ref{sec:NLOva}. For each
of the operators we stop at the order at which one expects the
contributions from the two-nucleon currents. For most of the
operators, this is ${\mathcal O}(q^{\nu_{\rm LO}+3})$; the exceptions
are the operators ${\cal O}_{2,q}^{(6)}$, ${\cal O}_{5,q}^{(7)}$,
${\cal O}_{6,q}^{(7)}$, for which the two-nucleon corrections arise at
${\mathcal O}(q^{\nu_{\rm LO}+1})$, and the operator ${\cal
  O}_{1,q}^{(6)}$, for which the corrections are of ${\mathcal
  O}(q^{\nu_{\rm LO}+2})$. Note that for ${\cal O}_{5,q}^{(7)}$,
${\cal O}_{6,q}^{(7)}$, and ${\cal O}_{1,q}^{(6)}$
the two-nucleon currents enter at the {\em same} order as the
subleading corrections. Partial results for the NLO nonrelativistic
reduction were derived in Ref.~\cite{Hoferichter:2016nvd}, where in
addition the two-nucleon corrections were considered.

Starting at subleading order there are terms that break Galilean
invariance. This is a consequence of the fact that the underlying theory
is Lorentz and not Galilean invariant~\cite{Heinonen:2012km}. These
corrections involve the average velocity of the nucleon before and
after the scattering event, $\vec v_a=(\vec k_1+\vec k_2)/(2 m_N)$,
and lead to ten new nonrelativistic operators listed in
Eqs. \eqref{eq:O1aO1a1}-\eqref{eq:O3a3}.

The operators that appear at subleading order in the nonrelativistic
reduction can have a qualitatively different structure from the ones
that arise at LO. For instance, the vector--vector current operator
$\Q_{1,q}^{(6)}= (\bar \chi \gamma_\mu \chi) (\bar q \gamma^\mu q)$
reduces at NLO to 
\begin{equation}
\begin{split}
\label{eq:Q1q6:NLO:simple}
\Q_{1,q}^{(6)}\to& F_1^{q/N}\op_1^N\Big(1+\cdots\Big)-\Big\{\big(F_1^{q/N}+F_2^{q/N}\big)\frac{\vec q^{\,\,2}}{m_\chi m_N}\op_4^N
-\big( F_1^{q/N}+F_2^{q/N}\big)\op_3^N
\\
&-\frac{m_N}{2 m_\chi} F_1^{q/N}\op_5^N-\frac{m_N}{ m_\chi}\Big(F_1^{q/N}+ F_2^{q/N}\Big)\op_6^N +\cdots\Big\}.
\end{split}
\end{equation}
At LO one thus has the number operator ${\mathcal O}_1^N=
\mathbb{1}_\chi \mathbb{1}_N$ and no spin dependence, while the
expansion to the subleading order gives in addition
velocity-suppressed couplings to spin through the operators ${\mathcal
  O}_4^N= \vec S_\chi \cdot \vec S_N$, $\op_{3,5}^N\sim \vec
S_{N,\chi}\cdot (\vec v_\perp \times \vec q)$, and $\op_6^N\sim (\vec
q\ncdot \vec S_N)(\vec q\ncdot \vec S_\chi)$. Such corrections could
have potentially important implications, if the LO expression leads to
incoherent, i.e., spin-dependent scattering, while at NLO there is a
contribution from the number operator ${\mathcal O}_1^N$. The latter
leads to an $A^2$-enhanced coherent scattering rate, where $A$ is the
mass number of the nucleus. For scattering on heavy nuclei with
$A\sim {\mathcal O}(100)$ the chirally subleading term can potentially
be the dominant contribution on nuclear scales.

There is only one operator, where this occurs, though. The
tensor-tensor operator, $\Q_{9,q}^{(7)}=m_q (\bar \chi \sigma^{\mu\nu}
\chi) (\bar q \sigma_{\mu\nu} q)$, leads at LO in the chiral expansion
to the spin-spin interaction, $\op_4^N = \vec S_\chi \cdot \vec S_N$.
At NLO, on the other hand, one also obtains a contribution of the form
$\sim \vec q^{\,\,2} \mathbb{1}_\chi \mathbb{1}_N$,
\begin{equation}
\begin{split}\label{eq:Q9q7:had:simple}
\Q_{9,q}^{(7)}\to&  8 F_{T,0}^{q/N}\op_4^N-\Big\{\frac{\vec
  q^{\,\,2}}{2 m_N m_\chi}\big(F_{T,0}^{q/N}-F_{T,1}^{q/N}\big)\op_1^N 
+\cdots \Big\}\,,
\end{split}
\end{equation} 
where we do not display the other $q^2$-suppressed terms. For heavy
nuclei the coherently enhanced contribution from $\op_1^N$ scales as
$A \vec q^{\,\,2}/(m_N m_\chi)\sim {\mathcal O}(1)$ and thus the
formally subleading contribution could, in principle, become important
in nuclear scattering. Inspection of this particular case, however,
shows that there is a relative numerical factor of 16 enhancing the
leading contribution. Furthermore the coherent ${\mathcal O}(q^2)$
term is suppressed by $1/m_N m_\chi$ and not simply by $1/m_N^2$,
further reducing its importance for heavy DM masses. As a result the
${\mathcal O}(q^2)$ terms are numerically unimportant also for the
tensor-tensor operator. In contrast, such coherent scattering is
important in $\mu\to e$ conversion, where the $m_\chi$ supression gets
replaced by $m_\mu$ \cite{Cirigliano:2017azj}.

A potential concern is that something similar, but with a less
favorable result for the numerical factors, could happen for some
other operator due to the uncalculated contributions from the
nonrelativistic expansion to even higher orders. However, one can
easily convince oneself that this is not the case by using the parity
properties of quark and DM bilinears. All the relativistic operators
in Eq.~\eqref{eq:lightDM:Lnf5} that are composed from parity-odd
bilinears necessarily involve the parity-odd spin operators for
single-nucleon currents at each order in the chiral expansion, because
one cannot form a parity-odd quantity from just two momenta -- the
incoming and the outgoing momentum
(cf. \eqref{eq:HDMETlimit:scalar:momenta}-\eqref{eq:axialtensorDM:expand:momenta}).
Such operators thus never lead to coherent scattering (the argument
above may need to be revisited for two-nucleon currents). This leaves
us with the operators composed from parity-even bilinears
only. Scalar--scalar operators and vector--vector operators lead to
coherent scattering already at LO, giving tensor--tensor operator as
the only left over possibility. The reduction of the tensor bilinear,
Eq. \eqref{eq:tensorDM:expand:momenta}, gives at LO $\sim
\epsilon^{\mu\nu\alpha\beta} v_{\alpha} S_{\beta}$, while at NLO one
also gets, among others, the combination $v^{[\mu}q^{\nu]}$. The
latter does not involve spin and leads to coherent
scattering. However, due to numerical prefactors, the latter
contribution is still subleading, as was shown above.

\section{Scalar dark matter}
\label{sec:scalarDM}
The above results are easily extended to the case of scalar
DM.\footnote{For operators and Wilson coefficients we adopt the same
  notation for scalar DM as for fermionic DM. No confusion should
  arise as this abuse of notation is restricted to this section and
  Appendix \ref{app:NR:scalar}.} For relativistic scalar DM, denoted
by $\varphi$, the effective interactions with the SM start at
dimension six,
\begin{equation}\label{eq:lightDM:Ln:scalar}
{\cal L}_\varphi=
\hat \C_{a}^{(6)} {\cal Q}_a^{(6)}+\cdots\,, 
\qquad {\rm where}\quad 
\hat \C_{a}^{(6)}=\frac{\C_{a}^{(6)}}{\Lambda^{2}}\,,  
\end{equation}
where ellipses denote higher dimension operators. The dimension-six
operators that couple DM to quarks and gluons are 
\begin{align}
\label{eq:dim6:Q1Q2:light:scalar}
{\cal Q}_{1,q}^{(6)} & = \big(\varphi^* i\overset{\leftrightarrow}{\partial_\mu} \varphi\big) (\bar q \gamma^\mu q)\,,
 &{\cal Q}_{2,q}^{(6)} &= \big(\varphi^* i\overset{\leftrightarrow}{\partial_\mu} \varphi\big)(\bar q \gamma^\mu \gamma_5 q)\,, 
 \\
 \label{eq:dim6:Q3Q4:light:scalar}
 {\cal Q}_{3,q}^{(6)} & = m_q (\varphi^* \varphi)( \bar q q)\,, 
&{\cal
  Q}_{4,q}^{(6)} &= m_q (\varphi^* \varphi)( \bar q i\gamma_5 q)\,,
\\
\label{eq:dim6:Q5Q6:light:scalar}
  {\cal Q}_5^{(6)} & = \frac{\alpha_s}{12\pi} (\varphi^* \varphi)
 G^{a\mu\nu}G_{\mu\nu}^a\,, 
 & {\cal Q}_6^{(6)} &= \frac{\alpha_s}{8\pi} (\varphi^* \varphi) G^{a\mu\nu}\widetilde G_{\mu\nu}^a\,.
\end{align}
while the coupling to photons are
\begin{align}
  \label{eq:dim6:Q7Q8:light:scalar} 
  {\cal Q}_{8}^{(6)} &=\frac{\alpha}{12\pi}(\varphi^*
\varphi) F^{\mu\nu}F_{\mu\nu}\,,
&  {\cal Q}_{9}^{(6)} &=\frac{\alpha}{8\pi}(\varphi^*
\varphi) F^{\mu\nu} \tilde F_{\mu\nu}. &~&
\end{align}
Here $\overset{\leftrightarrow}{\partial_\mu}$ is defined through
$\phi_1\overset{\leftrightarrow}{\partial_\mu} \phi_2=\phi_1
\partial_\mu \phi_2- (\partial_\mu \phi_1) \phi_2$, and $q=u,d,s$
again denote the light quarks. The strong coupling constant $\alpha_s$
is taken at $\mu\sim 1$ GeV, and $\alpha=e^2/4\pi$ the electromagnetic
fine structure constant. The operators ${\cal Q}_6^{(6)}$ and ${\cal
  Q}_9^{(6)}$ are CP-odd, while all the other operators are
CP-even. There are also the leptonic equivalents of the operators
${\cal Q}_{1,q}^{(6)},\ldots, {\cal Q}_{4,q}^{(6)}$, with $q\to \ell$.

At LO in chiral counting the operators coupling DM to quark and gluon
currents hadronize as 
\begin{align}
\begin{split}
\Q_{1q}^{(6)}\to& 2 F_1^{q/N} m_\varphi \op_1^N+{\mathcal O}(q^2)\,,
\end{split}
\\
\Q_{2q}^{(6)}\to &- 4 F_A^{q/N} m_\varphi \op_7^N+{\mathcal O}(q^3)\,,
\\
\Q_{3q}^{(6)}\to & F_S^{q/N} \op_1^N+{\mathcal O}(q^2)\,,
\\
\Q_{4q}^{(6)}\to & F_P^{q/N} \op_{10}^N+{\mathcal O}(q^3)\,,
\\
\Q_{5}^{(6)}\to & F_G \op_1^N+{\mathcal O}(q^2)\,,
\\
\Q_{6}^{(6)}\to & F_{\tilde G} \op_{10}^N+{\mathcal O}(q^3)\,.
\end{align}
The expressions valid to NLO in chiral counting are given in
Appendix~\ref{app:NR:scalar}. 

There are a number of qualitative differences between the cases of
fermionic and scalar DM. For instance, since scalar DM does not carry
a spin there is a much smaller set of operators that are generated in
the nonrelativistic limit. This greatly simplifies the
analysis. Furthermore, as opposed to the case of fermionic DM, there
are no cases where at LO in chiral counting one would obtain
incoherent scattering on nuclear spin, while at NLO in chiral counting
one would have coherent scattering.

\section{Examples}
\label{sec:examples}

In this section we discuss several numerical examples of DM direct
detection scattering. Most of the examples are for LO matching from
the EFT describing DM interacting with quarks and gluons onto a theory
that describes DM interacting with neutrons and protons in. At the end
of this section, we will also comment on the NLO corrections. The rate
${\cal R}$, i.e., the expected number of events per detector mass per
unit of time, is given by 
\begin{equation} \label{eq:dRdER} \frac{d{\cal
    R}}{dE_R}=\frac{\rho_\chi}{m_A\,m_\chi}\int_{v_{\rm
      min}}\frac{d\sigma}{dE_R} v f_{\oplus}(\vec v)d^3 \vec v\,,
\end{equation} 
where $E_R$ is the recoil energy of the nucleus, $m_A$ is the mass of
the nucleus, and $\rho_\chi$ is the local DM density. The integral is
over the DM velocity $v$ in the Earth's frame with a lower bound given
by $v_{\rm min}=\sqrt{m_A E_R/2}/\mu_{\chi A}$, where $\mu_{\chi
  A}=m_A m_\chi/(m_A+m_\chi)$ is the reduced mass of DM--nucleus
system. For the DM velocity distribution in the Earth's frame,
$f_\oplus(\vec v)$, we use the standard halo model, i.e., a
distribution that in the galactic frame takes the form of an isotropic
Maxwell-Boltzmann distribution with $v_0=254$ km/s (where
$v_0/\sqrt{2}$ is the width of the Gaussian), truncated at the escape
velocity $v_{\rm esc}=550$ km/s \cite{Olive:2016xmw}.

The DM-nucleus scattering cross section $d\sigma/dE_R$ in
Eq.~\eqref{eq:dRdER} is given by
\begin{equation}
\frac{d\sigma}{dE_R}=\frac{m_A}{2\pi v^2} \frac{1}{(2
  J_\chi+1)}\frac{1}{(2 J_A+1)}\sum_{\rm spins}|{\cal M}|_{\rm
  NR}^2\,. 
\end{equation}
The nonrelativistic matrix element squared is~\cite{Anand:2013yka} 
\begin{equation}\label{eq:Nucl:response:matrixel}
\begin{split}
\frac{1}{2 J_\chi+1}\,\frac{1}{2 J_A+1}\sum_{\rm spins}|{\cal M}|_{\rm  NR}^2=
&\frac{4\pi}{2J_A+1} \sum_{\tau=0,1}\sum_{\tau'=0,1}
\Big\{R_M^{\tau\tau'} W_M^{\tau\tau'}(q)+R_{\Sigma''}^{\tau\tau'} W_{\Sigma''}^{\tau\tau'}(q)
\\
+&R_{\Sigma'}^{\tau\tau'} W_{\Sigma'}^{\tau\tau'}(q)
+\frac{\vec q^{\,\,2}}{m_N^2}\Big[
R_{\Delta}^{\tau\tau'} W_{\Delta}^{\tau\tau'}(q)+R_{\Delta\Sigma'}^{\tau\tau'} W_{\Delta\Sigma'}^{\tau\tau'}(q)\Big]\Big\}, 
\end{split}
\end{equation}
where $J_\chi=1/2$ is the spin of DM in our examples and $J_A$ is the
spin of the target nucleus. The nuclear response function $W_i$ depend
on momentum exchange, $q \equiv |\vec q\,|$. The spin-independent
scattering is encoded in the response function $W_M$ which, for
instance, arises from the matrix element squared of the nuclear vector
current. In the long-wavelength limit, $q\to 0$, $W_M(0)$ simply
counts the number of nucleons in the nucleus giving coherently
enhanced scattering, $W_M(0)\propto A^2$. The response functions
$W_{\Sigma''}$ and $W_{\Sigma'}$ have the same long-wavelength limit
and measure the nucleon spin content of the nucleus. $W_\Delta$
measures the nucleon angular momentum content of the nucleus, while
$W_{\Delta \Sigma'}$ is the interference term. These functions roughly
scale as $W_M\sim {\mathcal O}(A^2)$, and $W_{\Sigma'}, W_{\Sigma''},
W_{\Delta}, W_{\Delta \Sigma'}\sim {\mathcal O}(1)$, where the actual
size depends on the particular nucleus and can differ significantly
from one nucleus to another. The prefactors $R_i$ encode the
dependence on the $c_i^N(q^2)$ coefficients, Eq.~\eqref{eq:LNR}, and
on kinematical factors. For instance, the coefficient of the
coherently enhanced term is
\begin{equation}
R_M^{\tau \tau'}=  c_{1}^{\tau} c_{1}^{\tau'}+
\frac{1}{4}\Big[\frac{\vec q^{\,\,2}}{m_N^2}c_{11}^{\tau}
  c_{11}^{\tau'}+\vec v_T^{\perp2} \Big(c_{8}^{\tau} c_{8}^{\tau'} +
  \frac{\vec q^{\,\,2}}{m_N^2}   c_{5}^{\tau} c_{5}^{\tau'}\Big)
  \Big]\,, 
\end{equation}
where $\vec v_T^\perp=\vec v -\vec q/(2 \mu_{\chi A})\sim
10^{-3}$. The sum in Eq.~\eqref{eq:Nucl:response:matrixel} is over
isospin values $\tau=0,1$ which are related to the proton and neutron
coefficients by $c_{i}^{0}=\big(c_{i}^{p}+c_{i}^{n}\big)/2,
c_{i}^{1}=\big(c_{i}^{p}-c_{i}^{n}\big)/2$. The remaining
$R_i^{\tau\tau'}$ can be found in \cite{Anand:2013yka}. Using these
expressions for $R_M^{\tau\tau'}$ together with our expressions for
the hadronization of the EFT operators,
Eqs.~\eqref{eq:Q15}-\eqref{eq:Q10q7:had:LO}, which give the
coefficients $c_i^\tau$ (see Appendix \ref{sec:anand}), we are now in
a position to obtain the rates in a DM direct detection experiment
assuming a particular interaction of DM with the visible sector.

\begin{figure}[t]\centering
\begin{minipage}{0.49\columnwidth}
	\hspace{1.2cm}$\C_{4,d}^{(6)}=-\C_{4,u}^{(6)}$\\
	\includegraphics[scale=1]{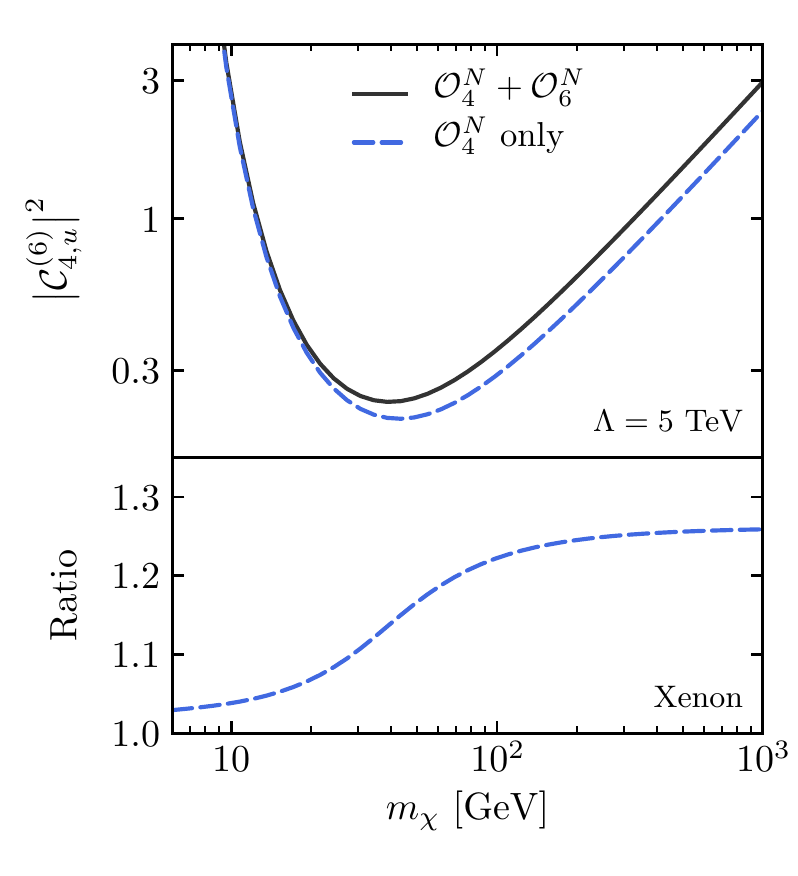}
\end{minipage}
\begin{minipage}{0.49\columnwidth}
	\hspace{1.2cm}$m_\chi = 100$ GeV\\
	\includegraphics[scale=1]{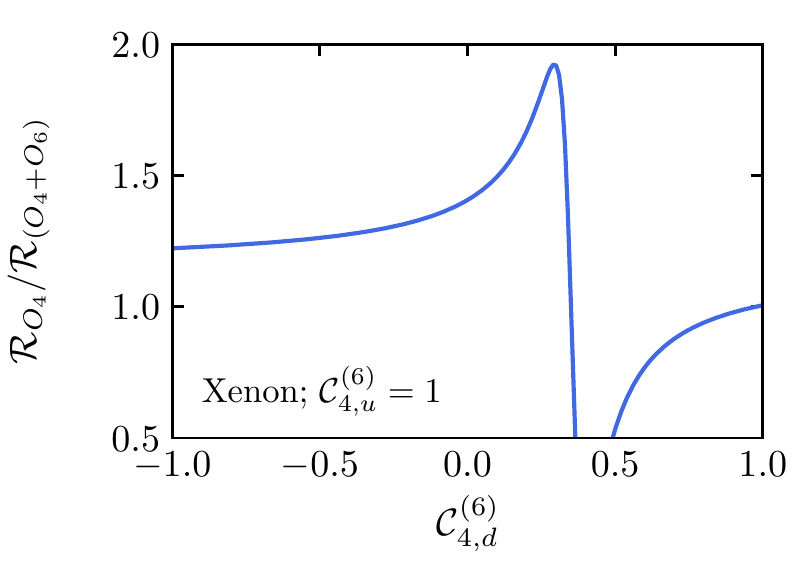}
\end{minipage}
	\caption{ {\it Left panel}: an illustration of Xe target
          bounds on the Wilson coefficients
          $\C_{4,u}^{(6)}=-\C_{4,d}^{(6)}$ for the interaction
          operator $(\bar \chi\gamma^\mu\gamma_5 \chi) (\bar
          q\gamma_\mu\gamma_5 q)$ assuming opposite couplings to the
          $u$ and $d$ quarks. The correct, chirally leading, treatment
          of the induced spin-dependent scattering with both
          $\op_4^N=\vec S_\chi \ncdot \vec S_N$ and $\op_6^N\propto
          (\vec S_\chi \ncdot \vec q)(\vec S_N \ncdot \vec q)$
          operators (black solid line) is compared to that of
          $\op_4^N$ only (blue dashed line). The ratio of the two is
          shown in the bottom plot. {\it Right panel}: the ratio of
          the $\mc{O}_4$ contribution to the rate over the total rate
          as a function of the Wilson coefficient $\C_{4,d}^{(6)}$ for
          a fixed value of $\C_{4,u}^{(6)}=1$, taking $m_\chi=100$
          GeV.  }
\label{fig:AxA}
\end{figure}

\begin{figure}\centering
	\begin{minipage}{0.49\columnwidth}
	\hspace{1.0cm}$\C_{4,u}^{(6)}=-\C_{4,d}^{(6)}=1$\\
	\includegraphics[width=\textwidth]{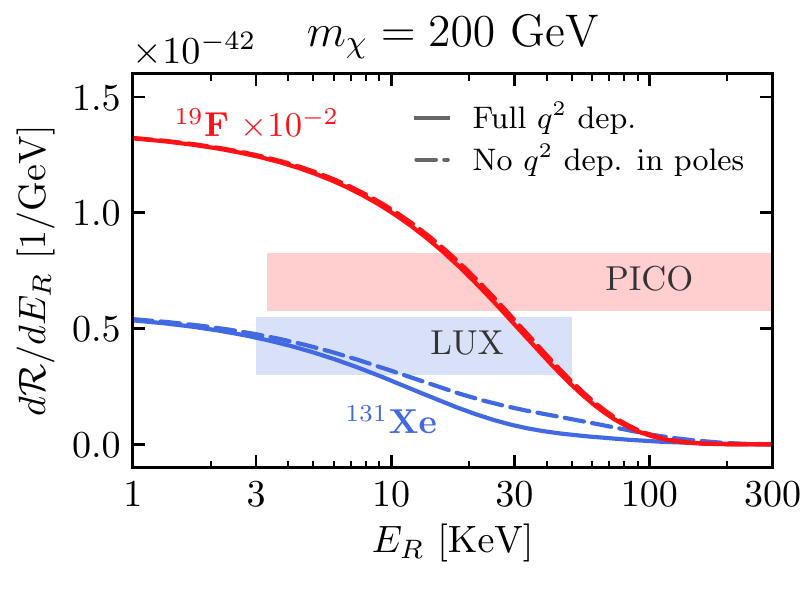}
\end{minipage}
\begin{minipage}{0.49\columnwidth}
	\hspace{1.0cm}$\C_{4}^{(7)}=1$\\
	\includegraphics[width=\textwidth]{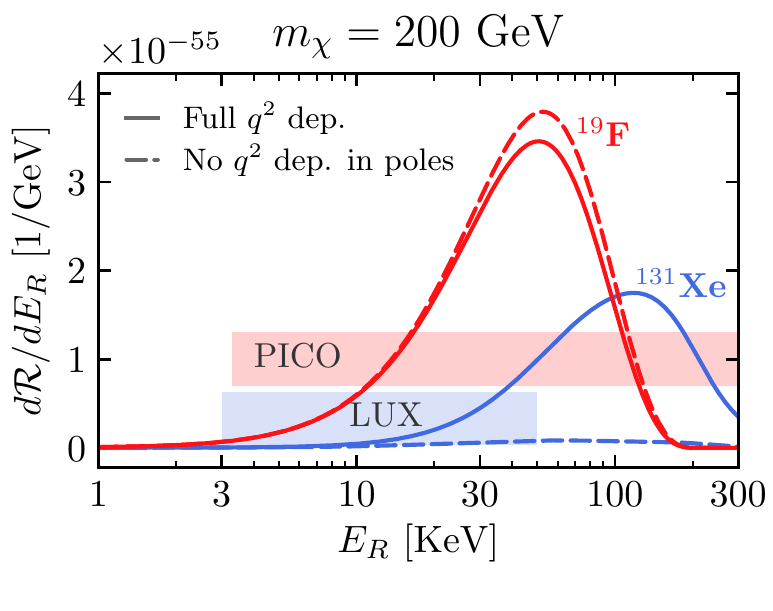}
\end{minipage}
	\caption{ The differential event rate, $d{\cal R}/dE_R$, as a
          function of the recoil energy, $E_R$, for scattering on
          Xenon (blue) and Fluorine (red) for $\Q_{4,q}^{(6)}$ and
          $\Q_4^{(7)}$ in the left and right panels respectively. In
          both panels, the solid curves include the full $q^2$
          dependence in the form factor $F_{\tilde G}(q^2)$ while the
          dashed lines include only the zero recoil limit, $F_{\tilde
            G}(0)$. The shaded regions depict the approximate ranges
          of experimental sensitivity for the LUX (blue) and PICO
          (red) experiments.  }
\label{fig:dRdQ-mx200}
\end{figure}

In the following, when we calculate the scattering rate and plot the
bound on the squared UV Wilson coefficients, we restrict the integral
over the recoil energy. To approximate the LUX sensitivity region we
integrate over $E_R\in[3,50]$ keV for Xenon~\cite{Akerib:2017kat}.  To
approximate PICO's~\cite{Amole:2017dex} sensitivity we integrated over
$E_R>3.3$ keV for Fluorine -- see Figs.~\ref{fig:dRdq}
and~\ref{fig:dRdQ-mx200}. To obtain total rates for scattering on
Xenon, we assume an exposure of 5000 kg$\cdot$yr which is
representative of the next generation two-phase liquid Xenon
detectors. Since Xenon has eight naturally occurring stable isotopes,
we sum over them weighted by their natural abundances.

\begin{figure}\centering
	\begin{minipage}{0.49\columnwidth}
		\includegraphics[scale=1]{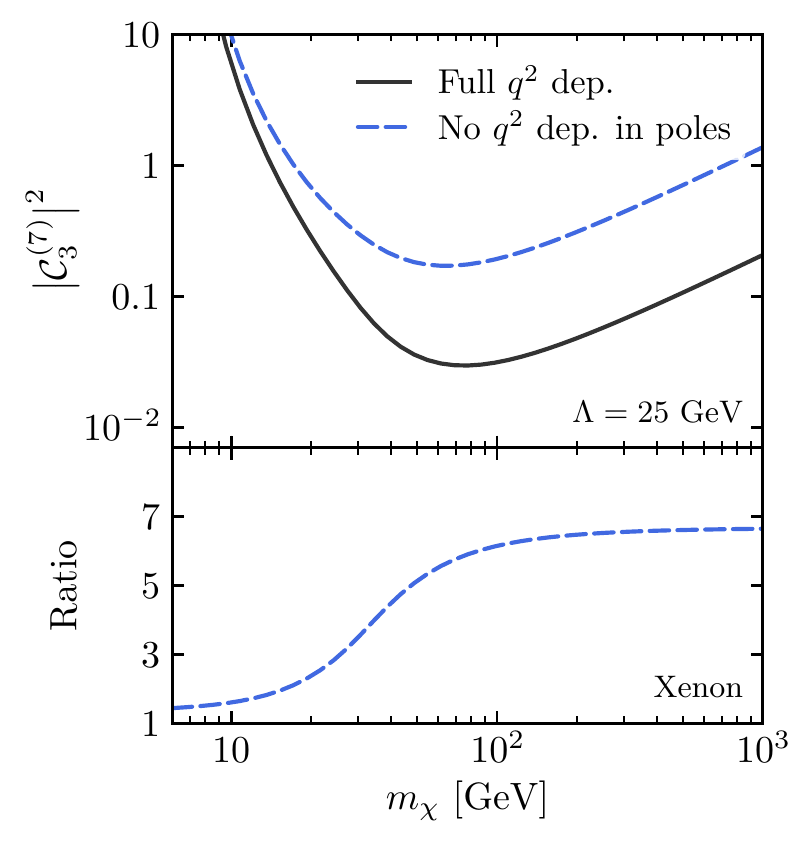}
	\end{minipage}
	\begin{minipage}{0.49\columnwidth}
		\includegraphics[scale=1]{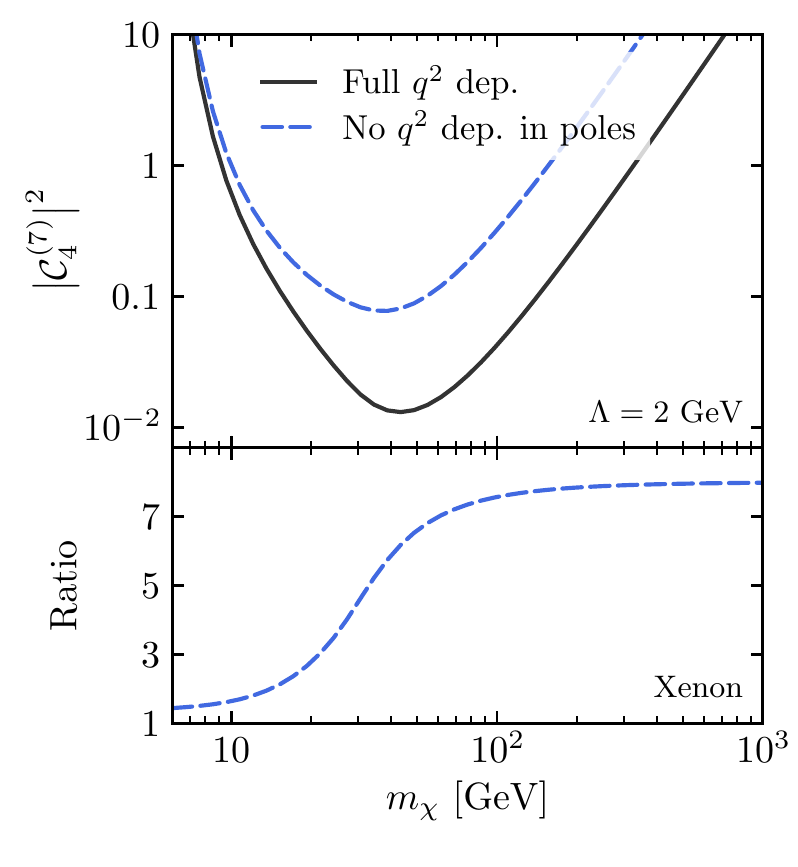}
	\end{minipage}
	\caption{
		Comparison between the bounds on the squared Wilson coefficients of the UV
		operators $\Q_3^{(7)}$ (left panel) and $\Q_{4}^{(7)}$ (right panel) for
		scattering on a Xenon target. The dashed and solid curves correspond to the
		bound with and without meson exchanges respectively. The lower plots show the
		ratio of the bounds without and with the inclusion of meson exchange.
	}
	\label{fig:SPxGGt}
\end{figure}

The first few examples, shown in Figs.~\ref{fig:AxA}, \ref{fig:dRdQ-mx200},
and \ref{fig:SPxGGt}, illustrate that one cannot always take the
long wavelength limit, $q\to 0$, in the calculation of DM scattering
rates when matching from ${\cal L}_\chi$ to ${\cal L}_{\rm NR}$. This
problem is well known for the description of DM scattering on {\em
  whole nuclei}, the effect described by the momentum dependence of
the nuclear response functions. For instance, a momentum exchange of
$q=100$ MeV already leads to decoherence and thereby reduces the
spin-independent nuclear form factor $W_M$ by $\sim 20\%$ ($\sim
60\%$) for scattering on Fluorine (Xenon). Our examples show a
different effect, namely that sometimes the momentum dependence cannot
be neglected even when considering the scattering on a {\em single}
neutron and/or proton. This effect is described by the momentum
dependence of the coefficients $c_i^{\tau\tau'}$. Since nucleons have
smaller spatial dimensions than nuclei, the effects of the momentum
dependence of $c_i^{\tau\tau'}$ are expected to be smaller than those
of the momentum dependence of $W_{i}^{\tau\tau'}$. However, because
the pseudoscalar hadronic currents contain pion poles, the corrections
due to non-zero momentum in the corresponding $c_i^{\tau\tau'}$ are of
${\mathcal O}(\vec q^{\,\,2}/m_\pi^2)$ and can be large.

The effect of such contributions for scattering on Xenon is shown in
Fig.~\ref{fig:AxA}. The chirally leading hadronization of the
axial-axial operator $(\bar \chi\gamma^\mu\gamma_5 \chi) (\bar
q\gamma_\mu\gamma_5 q)$ contains two nonrelativistic operators,
$\op_4^N=\vec S_\chi \ncdot \vec S_N$ and $\op_6^N\propto (\vec S_\chi
\ncdot \vec q\,)(\vec S_N \ncdot \vec q\,)$. The latter is momentum
suppressed but comes with a pion-pole enhanced coefficient, see
Eq.~\eqref{eq:Q46:schematic}, and thus gives an ${\mathcal O}(1)$
contribution to the scattering rate through interference with
$\op_4^N$. The left panel in Fig.~\ref{fig:AxA} shows a bound (solid
black line) on the relativistic Wilson coefficient $\C_{4,q}^{(6)}$
assuming equal and opposite couplings to the $u$ and $d$ quarks, and a
vanishing coupling to $s$ quarks.\footnote{In fact, we show a bound on
  $\big|\C_{4,q}^{(6)}\big|^2$ since this is directly proportional to
  the scattering rate.}  This is compared with the extraction of the
bound on $\C_{4,q}^{(6)}$ where the contribution of $\op_6^N$ is
neglected (dashed blue line). The two bounds coincide for small
$m_\chi$ since in that case the exchanged momenta are small which
parametrically suppresses the $\op_6^N$ contribution. The relative
difference then grows with $m_\chi$ up to $m_\chi\sim m_A$ (see lower
plot in Fig. ~\ref{fig:AxA} left), and is typically of
$\mc{O}(20\%-50\%)$, Fig.~\ref{fig:AxA} (right), confirming the
expectation from chiral counting that the correction is ${\mathcal
  O}(1)$ unless there are cancellations in one of the two
contributions. For instance, the $\op_4^N$ contribution is suppressed
for $\C_{4,d}^{(6)}\simeq \C_{4,u}^{(6)}/2$ and a DM mass $m_\chi=100$
GeV. Independent of the DM mass, however, the pion pole is completely
absent for $\C_{4,d}^{(6)}=\C_{4,u}^{(6)}$, and the $\op_6^N$
contribution to the scattering rate becomes negligible.

Furthermore, the contribution from $\op_6^N$ is expected to be
negligible for scattering on light nuclei since the exchanged momenta
are small, see Fig.  \ref{fig:dRdq}. We have explicitly checked this
for scattering on Fluorine, with the corresponding effect on $d{\cal
  R}/{dE_R}$ shown in Fig.  \ref{fig:dRdQ-mx200} (left) for
$m_\chi=200$ GeV. For scattering on ${}^{19}F$ the predictions with
(solid red line) and without $\op_6^N$ (dashed red line) essentially
coincide while for scattering on Xenon there is a large distortion of
the spectrum in the signal region for LUX.

The effect of pion exchange is even more pronounced if DM couples to
the visible sector through parity-odd gluonic operators, i.e., if the
operators in Eq.~\eqref{eq:dim7:Q3Q4:light} dominate. In
Fig. \ref{fig:SPxGGt}, we show the bounds on the Wilson coefficients
of the $\Q_3^{(7)}\propto \bar \chi \chi\, G \tilde G$ operator (left
panel), and of the operator $\Q_{4}^{(7)}\propto \bar \chi i \gamma_5
\chi \, G \tilde G$ (right panel). The corresponding nucleon form
factor has a schematic form
\begin{equation}
F_{\tilde G}(q)\sim \sum_i\frac{\Delta q_i}{m_{q_i}}+\delta m\,\frac{q^2}{m_{\pi,\eta}^2-q^2},
\end{equation}
where $\Delta q_i$ is the axial charge of quark $q_i$ and the $\delta
m$ coefficient is the size of isospin breaking for pion exchange and
the $SU(3)$-flavor breaking for eta meson exchange, see
Eq. \eqref{FGtilde:LO}. Note that isospin breaking is $\mathcal{O}(1)$
for the matrix element of the QCD anomaly term
$\alpha_s/(8\pi)\,G\tilde G$ while it is of $\mathcal{O}(10\%)$ for
all other matrix elements~\cite{Gross:1979ur}. The importance of
isospin-breaking but pion-pole enhanced contributions is reflected in
the DM scattering rates. The bounds on the Wilson coefficients
$\C_{3,4}^{(7)}$ in Fig. \ref{fig:SPxGGt}, obtained with the correct
full form factor dependence, are depicted with solid black lines. For
weak-scale DM masses they can be even up to an order of magnitude
stronger than the bounds obtained by only using the zero recoil form
factor, $F_{\tilde G}(0)$ (dashed blue lines). Ignoring the leading
$q^2$-dependence in $F_{\tilde G}$ also leads to a large distortion of
the shape in $d{\cal R}/dE_R$ as shown in Fig. \ref{fig:dRdQ-mx200}
(right) for the $\Q_{4}^{(7)}$ operator and $m_\chi=200$ GeV.  In this
case, there is a visible change in the shape of the differential rate
even for scattering on Fluorine, despite small momenta exchanges. The
effect is striking for the scattering on Xenon where momenta exchanges
are typically larger.  For the $\Q_{3}^{(7)}$ operator, the distortion
is slightly smaller, but otherwise comparable to the one shown.

\begin{figure}\centering
	\begin{minipage}{0.49\columnwidth}
	\hspace{1.2cm}$\C_{3,u}^{(6)}=\C_{3,d}^{(6)}=\C_{3,s}^{(6)}$\\
		\includegraphics[scale=1]{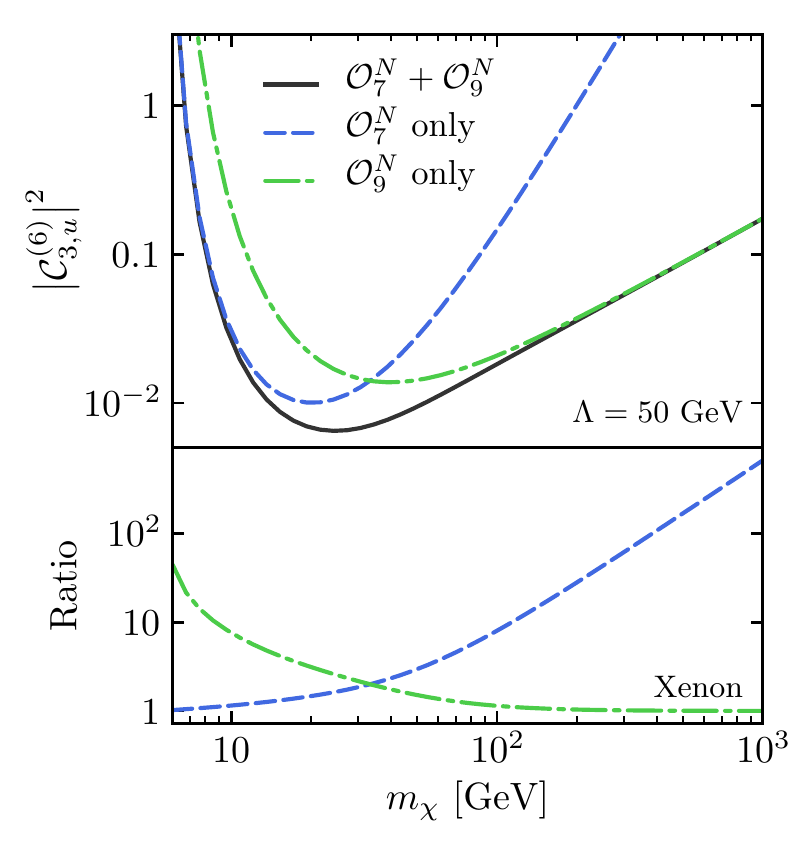}
	\end{minipage}
	\begin{minipage}{0.49\columnwidth}
		\hspace{1.2cm}$\C_{3,u}^{(6)}=\C_{3,d}^{(6)}=\C_{3,s}^{(6)}$\\
		\includegraphics[scale=1]{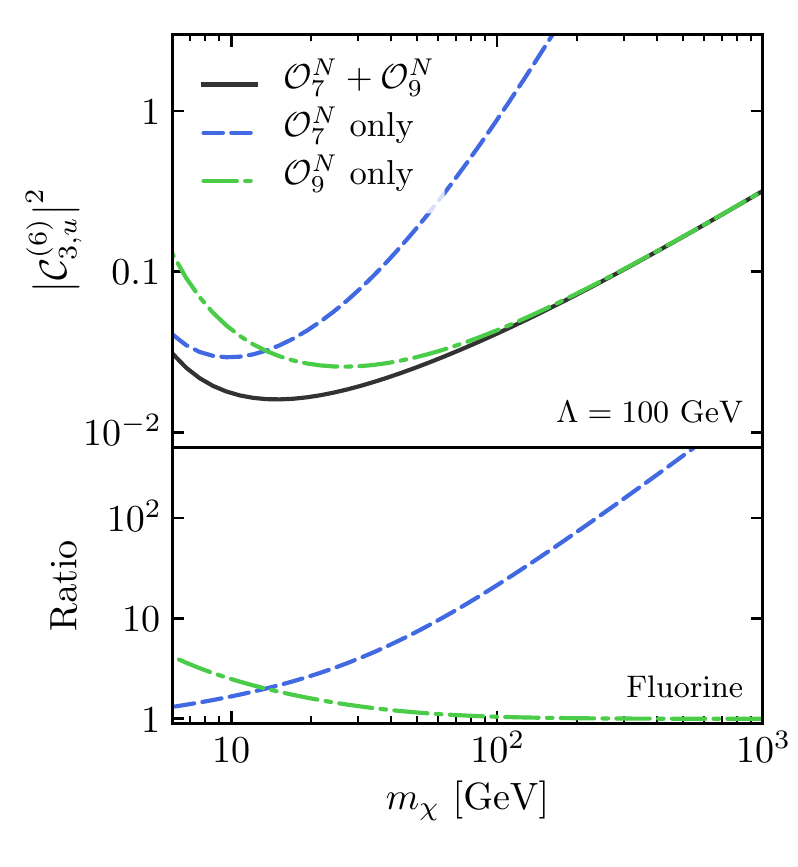}
	\end{minipage}
	\caption{
		The bounds on the squared Wilson coefficient of the $\mc{Q}_{3,q}^{(6)}=(\bar
		\chi \gamma^\mu \chi)(\bar q \gamma_\mu \gamma_5 q)$ operator from scattering
		on Xenon (left) and Fluorine (right), taking into account only $\op_7^N$
		(dashed blue line), only $\op_9^N$ operator (dot-dashed green line), and both
		(solid black). The coupling to all three light quarks are set equal to each
		other.
	}
	\label{fig:VxA}
\end{figure}

For the $\Q_{4,q}^{(6)}$ and $\Q_{4}^{(7)}$ operators discussed above
and shown for scattering on Xenon in
Figs.~\ref{fig:AxA}~and~\ref{fig:SPxGGt} respectively, the $\vec
q\,^2$ dependence in the meson poles is negligible for scattering on
Fluorine. To understand this it is useful to consider the differential
scattering rate as a function of the recoil energy. This is shown in
Fig.~\ref{fig:dRdQ-mx200} for a fixed DM mass of 200 GeV. For both
interactions, the $E_R$ spectra for Fluorine do not differ
significantly when the $\vec q\,^2$ dependence in the meson poles is
neglected since a given value of $E_R$ results in a momentum transfer
$\vec q\,^2/m_A$ that is smaller by an order of magnitude in Fluorine
than in Xenon.

\begin{figure}[t]\centering
	\begin{minipage}{0.49\columnwidth}
		\includegraphics[scale=1]{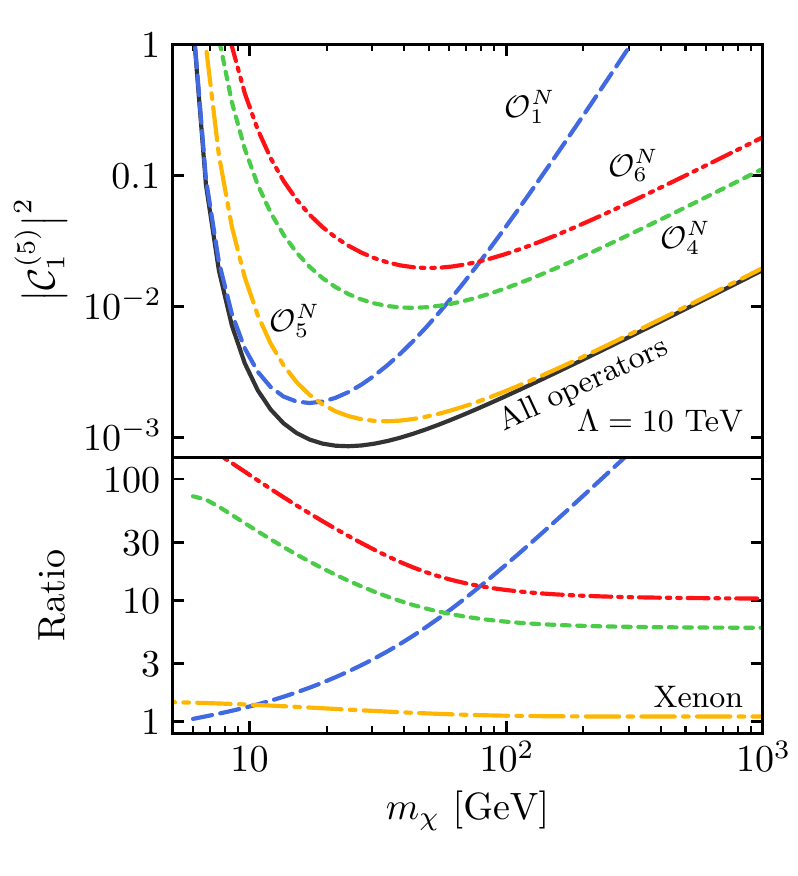}
	\end{minipage}
	\begin{minipage}{0.49\columnwidth}
		\includegraphics[scale=1]{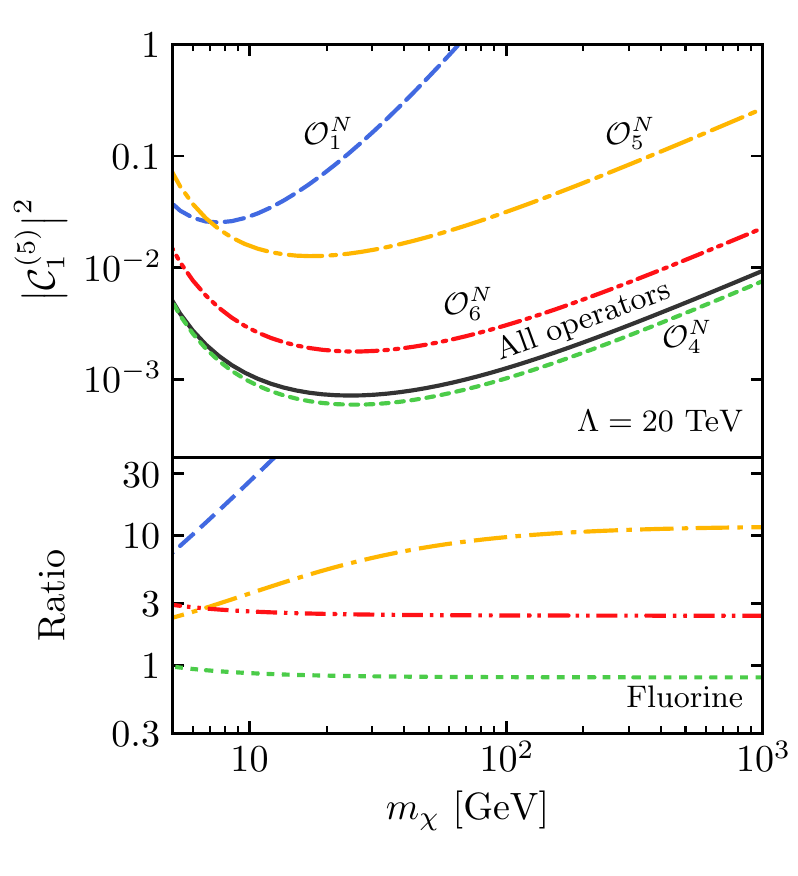}
	\end{minipage}
	\caption{
	The bound on the squared Wilson coefficient of the magnetic dipole
	operator $\Q_1^{(5)}$. The left (right) panel shows the scattering on Xenon
	(Fluorine). The EFT scale was fixed to 10 and 20 TeV for scattering on Xenon
	and Fluorine respectively. For both targets, the solid curve corresponds to the
	total rate while the dashed, dotted, dash-dotted and dash-double-dotted curves
	correspond to turning one non-relativistic operator at a time.
    }
    \label{fig:TxF}
\end{figure}

A qualitatively different example is given in Fig.~\ref{fig:VxA} which
shows the bounds on the Wilson coefficient $\C_{3,q}^{(6)}$ as
function of $m_\chi$ for scattering on Xenon and Fluorine.  The
vector-axial operator, $\Q_{3,q}^{(6)}=(\bar \chi \gamma^\mu\chi)(\bar
q\gamma_\mu \gamma_5 q)$, Eq.~(\ref{eq:dim6EW:Q3Q4:light}), matches
onto two non-relativistic operators, $\op_7^N\propto \vec S_N\cdot
\vec v_\perp$ and $\op_9^N\propto \vec S_\chi \cdot(\vec q\times \vec
S_N)$.
At leading order in chiral power counting, the hadronization of the
axial quark-current in $\Q_{3,q}^{(6)}$ is described by one form
factor at zero recoil , $F_A^{q/N}(0)$, see
Eq. \eqref{eq:LO:Q3q6}. This form factor is therefore a common
coefficient in the matching onto both $\op_7^N$ and $\op_9^N$.
Nevertheless, the contribution due to $\mcO_9^N$ is suppressed by an
additional power of the DM mass (i.e, two powers in the rate) and thus
becomes subleading for larger DM masses. Since the contributions are
correlated yet scale differently with $m_\chi$, it is crucial to
consider both non-relativistic operators when setting bounds from
direct detection experiments (see, e.g.,~\cite{Aprile:2017aas}).

The non-trivial interplay between different non-relativistic operators
can also be seen in the case of dipole interaction, $\Q_{1}^{(5)}$,
shown in Fig.~\ref{fig:TxF}. This operator matches onto four NR
operators $\op_1^N, \op_4^N, \op_5^N, \op_6^N$, see
Eq.~\eqref{eq:Q15}.  Out of these, two are coherently enhanced,
$\mcO_1^N=\mathbb{1}_\chi \mathbb{1}_N $ and $\mcO_5^N\propto \vec
S_\chi \cdot(\vec v_\perp\times \vec q)\mathbb{1}_N$. One expects
these two to dominate for heavier nuclei, as shown explicitly for
Xenon in Fig.~\ref{fig:TxF} (left).  The $\mcO_5^N$ operator is
enhanced by an explicit photon pole prefactor, $1/\vec q\,^2$, which
overcomes the velocity suppression and leads to its dominance over all
other contributions.  The contribution from the $\mcO_1^N$ operator,
on the other hand, is local and is suppressed for heavy DM by a
$1/m_\chi$ factor. Its contribution is, therefore, relevant only for
light DM.

For DM scattering on lighter nuclei, the situation is more
involved. The coherent enhancement is not as large and does not
overcome the velocity suppression in $\op_5^N$ even though it is
accompanied by the $1/\vec q\,^2$ enhancement. For $\mcO_{1}^N$, the
factor of $1/m_\chi$ still suppresses its contribution, particularly
for $m_\chi\gtrsim\mcO(10)$ GeV. For Fluorine the leading
contributions thus come from incoherent scattering due to the
spin-dependent $\mcO_4^N$ and $\mcO_6^N$ operators. Parametrically,
they scale in the same way (the $\vec q\,^2$ factor in $\mcO_6^N$ is
cancelled by the $1/\vec q\,^2$ in its Wilson
coefficient). Numerically, however, the contribution from $\op_4^N$ is
about three times larger. Furthermore, the contributions have opposite
signs and interfere destructively as can be seen in the right panel of
Fig.~\ref{fig:TxF}, with $\op_4^N$ giving a stronger bound than the
sum of all operators.

Finally, we turn our attention to the NLO corrections. The chiral
counting of the expansion in powers of $q^2$ is well motivated but
does not capture all effects. For instance, the NLO corrections in
chiral counting can become important if coherently enhanced operators
appear at NLO when there were none at LO. This is indeed the case for
the tensor operator $\Q_{9,q}^{(7)}$ where two coherently enhanced
operators, $\mcO_1^N$ and $\mcO_5^N$, appear at NLO in the expansion,
while at LO no coherently enhanced operators are present. However,
even for Xenon, the coherent enhancement is not enough to compensate
for the $\vec q\,^2/m_N m_\chi$ suppression accompanied by a relative
factor of $1/16$, and thus the resulting correction is of
$\mcO(5\%)$. A similar coherently enhanced contribution appears for
$\Q_{10,q}^{(7)}$ operator at ${\mathcal O}(q^4)$ and is thus
completely negligible.

\section{Conclusions }
\label{sec:conclusions}
In this article we derived the expressions for the matching of an EFT
for DM interacting with quarks and gluons, described by the effective
Lagrangian ${\cal L}_\chi$ in Eq. \eqref{eq:lightDM:Lnf5}, to an EFT
described by the Lagrangian ${\cal L}_{\rm NR}$ for nonrelativistic DM
interacting with nonrelativistic nucleons, Eq.~\eqref{eq:LNR}. The
latter is then used as an input to the description of DM interactions
with nuclei, described in terms of nuclear response functions. The
rationale underlying our work is the organization of different
contributions according to chiral power counting, i.e., in terms of an
expansion in $\vec q^{\,\,2}/\Lambda_{\rm ChEFT}^2$ and counting
$q\sim m_\pi$. Within this framework one can make the following
observations: (i) for LO expressions one needs nonrelativistic
operators with up to two derivatives, since they can be enhanced by
pion poles giving a contribution of the order of $\vec
q^{\,\,2}/(m_\pi^2+\vec q^{\,\,2})\sim {\mathcal O}(1)$; (ii) not all
of the nonrelativistic operators $\op_i^N$ with two derivatives are
generated when starting from an EFT for DM interacting with quarks and
gluons; (iii) a single relativistic operator $\Q_i^{(d)}$ can generate
several nonrelativistic operators $\op_i^N$ with momentum-dependent
coefficients already at LO; (iv) interactions of DM with two-nucleon
currents are chirally suppressed (barring cancellations of LO terms),
justifying our treatment of DM interacting with only single-nucleon
currents.

We worked to next-to-leading order in the chiral expansion, but also
discussed separately the expressions for the leading-order
matching. At LO the scattering of DM on nucleons only depends on the
DM spin $\vec S_\chi$, the nucleon spin $\vec S_N$, the momentum
exchange $\vec q$, and the averaged relative velocity between DM and
nucleon before and after scattering, $\vec v_\perp$. All these
quantities are Galilean invariant. At NLO in chiral counting the
expressions depend in addition on the averaged velocity of nucleon
before and after scattering, $\vec v_a$. This dependence on Galilean
non-invariant quantities such as $\vec v_a$ is expected, since the
underlying theory is Lorentz and not Galilean invariant. Because of
the dependence on $\vec v_a$ the NLO expressions require an expanded
nonrelativistic operator basis, with the new operators listed in
Appendix~\ref{sec:NLOva}.

Numerically the NLO corrections are always small, at the level of
${\mathcal O}(\vec q^{\,\,2}/m_N^2)$ or a few percent, unless one fine
tunes the cancellation of LO expressions. This result is nontrivial
for the partonic tensor-tensor operator $Q_{9,q}^{(7)} = m_q (\bar
\chi \sigma^{\mu\nu} \chi) (\bar q \sigma_{\mu\nu} q)$,
since in that case the LO term is spin-dependent, while the NLO
corrections contain a spin-independent contribution that is coherently
enhanced. In principle this could compete with the LO term. However,
due to fortuitous numerical factors, it remains subleading.

While our results were obtained by assuming that the mediators between
the DM and the visible sector are heavy, with masses above several
hundred MeV, the formalism can be easily changed to accommodate
lighter mediators. In this case the mediators cannot be integrated
out, but lead to an additional momentum dependence of the coefficients
in the nonrelativistic Lagrangian ${\cal L}_{\rm NR}$,
Eq. \eqref{eq:LNR}, and potentially to a modified counting of chirally
leading and subleading terms. The details of the latter would depend
on the specifics of the underlying DM theory.

As a side-result, our expressions show that from the particle physics
point of view it is more natural to interpret the results of direct
detection experiments in terms of an EFT where DM interacts with
quarks and gluons, Eq. \eqref{eq:lightDM:Lnf5}. The reason is that several
of the partonic operators in ${\cal L}_\chi$ match to more than one
nonrelativistic operator already at leading order in chiral
counting. In such cases it is then hard to justify singling out just
one nonrelativistic operator in the analysis of direct detection
experimental results.

The situation becomes even more complicated if the partonic operator
matches onto several nuclear operators with different momentum
dependence, since in the experiments one integrates over a range of
momenta. A cautionary example of wider phenomenological interest is
the case of the axial-axial partonic operator, $Q_{4,q}^{(6)}=(\bar
\chi\gamma_\mu\gamma_5 \chi)(\bar q \gamma^\mu \gamma_5 q)$, which
induces spin-dependent scattering. At leading chiral order this is
described by a combination of the $\op_4^N=\vec S_\chi \cdot \vec S_N$
and $\op_6^N\sim \big(\vec S_\chi \ncdot \vec q\,\big)\big(\vec S_N
\ncdot \vec q \,\big)$ nonrelativistic operators. Naively the latter
is momentum suppressed. We find that this is true for DM scattering on
light nuclei, such as Fluorine, where the contribution from $\op_6^N$
is in fact unimportant, since the momenta exchanges are in this case
small, $q\ll m_\pi$. However, for DM scattering on heavy nuclei, such
as Xenon, the $\op_6^N$ operator does give an ${\mathcal O}(1)$
correction due to its enhancement by a pion pole, in line with the
expectations from chiral counting. Thus, in general both contributions
from $\op_4^N$ and $\op_6^N$ need to be kept.

The flip side of the above discussion is the question: are there
models of DM where only $\op_4^N$ or only $\op_6^N$ operator is
generated? For these two operators the answer is yes. At leading
chiral order the partonic operator $\Q_{9,q}^{(7)}= m_q (\bar \chi
\sigma^{\mu\nu} \chi) (\bar q \sigma_{\mu\nu} q)$ only generates
$\op_4^N$, while the partonic operators $\Q_4^{(7)}\sim (\bar \chi i
\gamma_5 \chi) G \widetilde G $, $\Q_{8,q}^{(7)}= m_q (\bar \chi i
\gamma_5 \chi)(\bar q i \gamma_5 q)$ only induce the operator
$\op_6^N$. But the same is not true in general.  For a number of other
nonrelativistic operators -- $\op_7^N,\op_8^N, \op_9^N$ and
$\op_{12}^N$ -- there is no partonic level operator that would induce
just one of these.  All of them are always accompanied by other
nonrelativistic operators when matching from ${\cal L}_\chi$ to ${\cal
  L}_{\rm NR}$.  For these nonrelativistic operators switching on just
one operator at the time when analysing direct detection data thus
does not make much sense from the microscopic point of view.
Furthermore, the nonrelativistic operators $\op_2^N$, $\op_3^N$,
$\op_{13}^N$, $\op_{14}^N$, $\op_{15}^N$, $\op_{2b}^N$ are never
generated as leading operators when starting from a UV theory of
DM. They enter only as subleading corrections in the scattering rates,
and can always be neglected (as can the other nine nonrelativistic
operators listed in Appendix \ref{sec:NLOva} that have already never
been considered).

In conclusion, we advocate the use of partonic level EFT basis
Eqs.~\eqref{eq:dim5:nf5:Q1Q2:light}-\eqref{eq:dim7EW:Q9Q10:light} as a
phenomenologically consistent way of interpreting direct detection
data. Including all the variations due to quark flavor assignments
there are 34 operators in total, which is not much more than the 28
nonrelativistic operators used at present. Moreover, using the
partonic level EFT also has the added benefit of providing a simple
connection with the use of EFT in collider searches for dark matter,
via straight-forward renormalization-group evolution.

{\bf Acknowledgements} We thank Christian Bauer, Eugenio Del Nobile,
Ulrich Haisch, Matthew McCullough, Paolo Panci, Mikhail Solon, and
Alessandro Strumia for useful discussions. FB is supported by the
Science and Technology Facilities Council (STFC). JZ is supported in
part by the U.S. National Science Foundation under CAREER Grant
PHY-1151392 and by the DOE grant de-sc0011784. BG is supported in part
by the U.S. Department of Energy under grant DE-SC0009919.
  
\appendix
\section{Values of the nucleon form factors}
\label{App:form:factors}

Below we give the values for the form factors $F_{i}^{p/q}$ for
proton external states, while the corresponding values for neutrons
are obtained through exchange of $p\to n$, $u\leftrightarrow d$.

\subsection{Vector current} 
\label{app:sec:vector:current}
The general matrix element of the vector
current~\eqref{vec:form:factor} is parameterized by two sets of form
factors $F_{1}^{q/N}(q^2)$ and $F_{2}^{q/N}(q^2)$. For the LO
expressions we only need their values evaluated at $q^2=0$, while for
the subleading expression~\eqref{eq:Q1q6:NLO} we also need
$F_{1}^{\prime \, q/N}(0)$.

At zero momentum exchange the vector currents count the number of
valence quarks in the nucleon. Hence, the normalization of the Dirac
form factors for the proton is
\begin{equation}\label{eq:F1:num}
F_1^{u/p}(0)=2, \qquad F_1^{d/p}(0)=1, \qquad F_1^{s/p}(0)=0.
\end{equation}
The Pauli form factors $F_2^{q/N}(0)$ describe the contributions of
quarks to the anomalous magnetic moments of the nucleons,
\begin{equation}
\begin{split}
  a_p & = \frac{2}{3} F_2^{u/p} (0) - \frac{1}{3}
  F_2^{d/p} (0) - \frac{1}{3} F_2^{s/p} (0) \approx 1.793 \,,\\
  a_n & = \frac{2}{3} F_2^{u/n} (0) - \frac{1}{3}
  F_2^{d/n} (0) - \frac{1}{3} F_2^{s/n} (0) \approx -1.913 \,.
\end{split}
\end{equation}
Using the strange magnetic moment~\cite{Sufian:2016pex} (see
also~\cite{Green:2015wqa}) 
\begin{equation}\label{eq:F2s:num}
F_2^{s/p}(0)= -0.064(17)\,,
\end{equation}
one gets, using isospin symmetry, 
\begin{align}\label{eq:F2:num}
F_2^{u/p}(0)&=2 a_p+a_n+F_2^{s/p}(0)=1.609(17)\,,
\\
\label{eq:F2d:num}
F_2^{d/p}(0)&=2 a_n+a_p+F_2^{s/p}(0)=-2.097(17)\,.
\end{align}
For the slope of $F_1^{q/N}(q^2)$ at $q^2=0$ one obtains~\cite{Hill:2014yxa}
\begin{align}
F_1^{\prime \, u/p}(0)& = \frac{1}{6}\big(2 \big[r_E^{p} \big]^2 + 
  \big[r_E^{n} \big]^2 +r_s^2\big)-\frac{1}{4m_N^2}\big(2
  a_p + a_n)=5.57(9){\rm~GeV}^{-2}\,, 
  \\
F_1^{\prime \, d/p}(0) & = \frac{1}{6}\big(\big[r_E^{p} \big]^2 +
  2 \big[r_E^{n} \big]^2 +r_s^2\big)-\frac{1}{4m_N^2}\big(
  a_p + 2a_n)=2.84(5){\rm~GeV}^{-2}\,, \label{eq:F1pdp0:num}
  \\
  F_1^{\prime \, s/p}(0) & = \frac{1}{6} r_s^2=-0.018(9){\rm~GeV}^{-2}\,,
\end{align}
using the values $\big[ r_E^{p} \big]^2 =
0.7658(107)\,$fm$^2$~\cite{Mohr:2015ccw, Olive:2016xmw}, 
$\big[ r_E^{n} \big]^2 = -0.1161(22)\,$fm$^2$~\cite{Olive:2016xmw},
and $r_s^2 = -0.0043(21)\,$fm$^2$~\cite{Sufian:2016pex}.

Above we used the definitions for the proton and neutron matrix
elements of the electromagnetic current, 
\begin{equation}
  \langle N'| J_\text{em}^\mu |N\rangle = \bar u_N' \Big[
    F_1^{N}(q^2) \gamma^\mu + \frac{i}{2m_N} F_2^{N}(q^2)
    \sigma^{\mu\nu} q_\nu \Big] u_N\,,\qquad N=p,n\,,
\end{equation}
where $J_\text{em}^\mu=\big(2 \bar u \gamma^\mu u- \bar d \gamma^\mu
d-\bar s \gamma^\mu s)/3$. The Sachs electric and magnetic form
factors are related to the Dirac and Pauli form factors, $F_1^N$ and
$F_2^N$, through \cite{Ernst:1960zza} (see also, e.g.,
\cite{Hill:2010yb})
\begin{equation}
G_E^N(q^2)=F_1^N(q^2)+\frac{q^2}{4 m_N^2} F_2^N(q^2)\,, \quad {\rm and}\quad
G_M^N(q^2)=F_1^N(q^2)+F_2^N(q^2)\,.
\end{equation}
 At zero recoil one has for the electric form factor, $G_E^p(0)=1,
 G_E^n(0)=0$, while the magnetic form factor at zero recoil gives~\cite{Olive:2016xmw}, 
 \begin{equation}
 G_M^p(0)=\mu_p\simeq 2.793, \qquad G_M^n(0)=\mu_n\simeq-1.913,
 \end{equation}
  i.e., the proton
 and neutron magnetic moments in units of nuclear magnetons $\hat\mu_N
 =e/(2m_N)$. The anomalous magnetic moments are $F_2^p(0)=a_p$,
 $F_2^n(0)=a_n$. The charge radii of the proton and neutron are
 defined through 
\begin{equation}
G_E^N(q^2)=G_E^N(0)+\frac{1}{6}\big[r_E^N]^2 q^2+\cdots\,.
\end{equation}

\subsection{Axial vector current} 

The matrix element of the axial-vector
current~\eqref{axial:form:factor} is parametrized by two sets of form
factors, $F_{A}^{q/N}(q^2)$ and $F_{P'}^{q/N}(q^2)$. For the LO
expressions we only need $F_{A}^{q/N}(0)$ and the light meson pole
parts of $F_{P'}^{q/N}(q^2)$,
\begin{equation}
\label{eq:F_P':LO}
F_{P'}^{q/N}(q^2)=\frac{m_N^2}{m_\pi^2-q^2} a_{P',\pi}^{q/N}+\frac{m_N^2}{m_\eta^2-q^2} a_{P',\eta}^{q/N}+\cdots\,.
\end{equation} 
The axial vector form factors $F_A^{q/N}$ at zero momentum transfer
are obtained from the matrix elements $2 m_p s^\mu \Delta q_p =
\langle p |\bar q \gamma^\mu \gamma_5 q|p\rangle_Q$, where $|p\rangle$
and $\langle p|$ denote proton states at rest.  Moreover, $s^\mu$ is
the proton's polarization vector such that $s^2=-1, s\cdot k_{p}=0$,
where $k_p^\mu=m_p(1,0,0,0)$ is the proton four-momentum, and the
matrix element is evaluated at scale $Q$. Consequently we find
\begin{equation}\label{eq:FA:def}
F_A^{q/p}(0)=\Delta q_p,
\end{equation}
while for the residua of the pion- and eta-pole contributions to
$F_{P'}^{q/N}$ we have 
\begin{align}
\label{eq:aP'pi}
a_{P',\pi}^{u/p}&=-a_{P',\pi}^{d/p}=2 g_A\,, \qquad a_{P',\pi}^{s/p}=0\,,
\\
\label{eq:aP'eta}
a_{P',\eta}^{u/p}&=a_{P',\eta}^{d/p}=-\frac{1}{2}a_{P',\eta}^{s/p}=\frac{2}{3}\big(\Delta
u_p+\Delta d_p-2 \Delta s_p\big) \,.
\end{align}
As always, the coefficients for the neutrons are obtained through a
replacement $p\to n, u\leftrightarrow d$ (no change is implied for
$g_A$).  We work in the isospin limit, so that 
\begin{equation}
\label{eq:Deltau:def}
\Delta u\equiv \Delta u_p=\Delta d_n, \qquad
\Delta d\equiv \Delta d_p=\Delta u_n, \qquad
\Delta s\equiv \Delta s_p=\Delta s_n.
\end{equation}
The isovector combination is determined precisely from nuclear $\beta$
decay~\cite{Olive:2016xmw}, 
\begin{equation}\label{eq:App:u-d}
\Delta u-\Delta d=g_A=1.2723(23).
\end{equation}
In the $\overline{\rm MS}$ scheme at $Q=2$~GeV the averages of
lattice QCD results give $\Delta u+\Delta
d=0.521(53)$~\cite{diCortona:2015ldu},
$\Delta s=-0.031(5)$ 
(averaging over
\cite{QCDSF:2011aa, Engelhardt:2012gd,
  Bhattacharya:2015gma,Alexandrou:2017hac} 
and inflating the errors
in~\cite{Bhattacharya:2015gma} by a factor of 2 because no continuum
extrapolation was performed). Combining with Eq.~\eqref{eq:App:u-d}
this gives~\cite{diCortona:2015ldu}
\begin{equation}
\label{eq:Deltaq:values}
\Delta u=0.897(27), \qquad \Delta d=-0.376(27), \qquad \Delta s=-0.031(5),
\end{equation}
all at the scale $Q=2\,$GeV.  The experiments give $\Delta
u=0.843(12)$, $\Delta d=-0.427(12)$  \cite{Alexandrou:2017hac},
  in good agreement with the lattice QCD, and a
somewhat larger value for the $s$-quark, $\Delta s=-0.084\pm0.017$,
averaging over HERMES
\cite{Airapetian:2006vy} and COMPASS \cite{Alexakhin:2006oza} results
(see also axion review in \cite{Olive:2016xmw}). 
  Note that, while the matrix elements
$\Delta q$ are scale dependent, the non-isosinglet combinations
$\Delta u-\Delta d$ and $\Delta u +\Delta d- 2 \Delta s$ are scale
independent, since they are protected by non-anomalous Ward
identities.

The derivative of the axial form factor at zero recoil is well known
for the $u-d$ current. Using the dipole ansatz~\cite{Meyer:2016oeg}
gives $F_A'(0)/F_A(0)=2/m_A^2$, with $m_A$ the
appropriate dipole mass. A global average over
experimental~\cite{Bernard:2001rs,
  AguilarArevalo:2010zc} and lattice \cite{Alexandrou:2017hac,
  Rajan:2017lxk} 
gives for the $u-d$ current dipole mass $m_A^{u-d}=1.064(29)$GeV,
 rescaling the combined error following the PDG
prescription (the $z$-expansion analysis leads to larger error
estimates, corresponding to
$m_A^{u-d}=1.01(24)$GeV~\cite{Meyer:2016oeg}),
while for the
$u+d$ current one has $m_A^{u+d}=1.64(14)$GeV
\cite{Alexandrou:2017hac,Bratt:2010jn} and for
the strange-quark current, $m_A^s=0.82(21)$ GeV
\cite{Alexandrou:2017hac}. This gives 
\begin{equation}
F_A^{u/p}{}'(0)=1.32(7)~{\rm GeV}^{-2}\,,\qquad F_A^{d/p}{}'(0)=-0.93(7)~{\rm GeV}^{-2}\,,
\end{equation}
or in terms of normalized derivatives 
\begin{equation}
\frac{F_A^{u/p}{}'(0)}{F_A^{u/p}(0)}=1.47(8)~{\rm GeV}^{-2}\,,\qquad
\frac{F_A^{d/p}{}'(0)}{F_A^{d/p}(0)}=2.47(22)~{\rm GeV}^{-2}\,,
\end{equation}
while for the strange quark
\begin{equation}
\frac{F_A^{s/p}{}'(0)}{F_A^{s/p}(0)}=\big(3.0\pm1.5\big)~{\rm GeV}^{-2}\,.
\end{equation}

At NLO $F_{P'}^{q/N}(q^2)$ needs to be expanded to
\begin{equation}
\label{eq:F_P':NLO}
F_{P'}^{q/N}(q^2)=\frac{m_N^2}{m_\pi^2-q^2}
a_{P',\pi}^{q/N}+\frac{m_N^2}{m_\eta^2-q^2}
a_{P',\eta}^{q/N}+b_{P'}^{q/N}+\cdots\,. 
\end{equation} 
At NLO the residua of the poles change by corrections of ${\mathcal
  O}\big(m_{\pi,\eta}^2/(4\pi f_\pi^2)^2\big)\approx0.01-0.05$. For
instance, for the $u-d$ current one has at NLO in HBChPT
\cite{Bernard:1998gv},
\begin{equation}\label{eq:FP':def}
F_{P'}^{(u-d)/p}=\frac{4m_N^2}{m_\pi^2-q^2}\biggr[g_A-\frac{2 m_\pi^2
    \tilde B_2}{(4\pi f_\pi)^2}\biggr]-\frac{2}{3} g_A m_N^2 r_A^2\,,
\end{equation}
where $\tilde B_2\approx -1.0\pm0.5$  is the
HBChPT low energy constant, while $r_A^2=6 F_{A}'(0)/F_{A}$. 
The constant term $b_{P'}$ is, therefore, for the $u-d$ current given
by 
\begin{equation}\label{bP'u-d}
b_{P'}^{(u-d)/p}=
-4 g_A m_N^2 \frac{F_{A}^{(u-d)/p}{}'(0)}{F_{A}^{(u-d)/p}(0)}\,.
\end{equation}
Assuming that the relation \eqref{bP'u-d} is valid for each quark
flavor separately, i.e., neglecting the anomaly contribution to
$b_{P'}^{q/p}$, gives
\begin{equation}\label{bP'num}
b_{P'}^{u/p}\approx-4.65(25)\,,\qquad b_{P'}^{d/p}\approx 3.28(25)\,, \qquad b_{P'}^{s/p}\approx (-11\pm6)\Delta s\,.
\end{equation}
as well as $b_{P'}^{s/p} \approx 0.32(18)$.
In our numerical analysis we estimated the importance of NLO
corrections by keeping $a_{P',\pi}^{q/N}$, $a_{P',\eta}^{q/N}$ at
their LO values, while setting $b_{P'}^{q/N}$ to the values in
\eqref{bP'num}. Note that these are a small correction to the LO
expression when the pion pole is present, but can be important when
this is not the case.

\subsection{Scalar current}
The scalar form factors $F_S^{q/N}$, Eq. \eqref{scalar:form:factor},
evaluated at $q^2=0$ are conventionally referred to as nuclear sigma
terms, 
\begin{equation}\label{eq:FS:def}
F_S^{q/N}(0)= \sigma_q^N\,,
\end{equation}
where $ \sigma_q^N \bar u_N u_N=\langle N |m_q\bar q q|N \rangle$,
$|N\rangle$ and $\langle N|$ represent the nucleon states at rest.
  Another common notation
is $\sigma_q^N=m_Nf_{Tq}^{N}$. Taking the naive average of the most
recent lattice QCD determinations~\cite{Junnarkar:2013ac,
  Yang:2015uis, Durr:2015dna}, we find 
  \begin{equation}
  \label{eq:values:sigmas}
  \sigma_s^p=\sigma_s^n=(41.3\pm 7.7){\rm~MeV}\,.
  \end{equation}
The matrix elements of the $u$ and $d$ quarks are related to the
pion-nucleon sigma term, defined as $\sigma_{\pi N} = \langle N | \bar
m (\bar u u + \bar d d) | N \rangle$, where $\bar m = (m_u +
m_d)/2$. A Heavy Baryon Chiral Perturbation Theory analysis of the
$\pi N$ scattering data gives $\sigma_{\pi N} =
59(7)$\,MeV~\cite{Alarcon:2011zs}, and a fit of $\pi N$ scattering
data to a representation based on Roy-Steiner equations gives
$\sigma_{\pi N}=58(5)$~MeV~\cite{RuizdeElvira:2017stg}. A more precise
determination is obtained from pionic atoms, $\sigma_{\pi
  N}=(59.1\pm3.5)$~MeV~\cite{Hoferichter:2015dsa}.  These are in
agreement with $\sigma_{\pi N}=52(3)(8)$ MeV obtained from a fit to
world lattice $N_f=2+1$ QCD data at the
time~\cite{Alvarez-Ruso:2014sma}. Including, however, both $\Delta
(1232)$ and finite spacing in the fit shifted the central value to
$\sigma_{\pi N}=44$ MeV.  More recent lattice QCD determinations
prefer an even slightly lower value, $\sigma_{\pi N}=38(2)$ MeV (the
average of results in~\cite{Durr:2015dna, Abdel-Rehim:2016won,
  Yang:2015uis}, see also remarks in~\cite{Hoferichter:2016ocj}). We
thus use a rather conservative estimate $\sigma_{\pi N}=(50\pm15)$
MeV. Using the expressions in~\cite{Crivellin:2013ipa} this gives
\begin{equation}
\label{eq:values:sigmaqN}
\begin{split}
\sigma_u^p=(17\pm 5){\rm~MeV}\,, \qquad \sigma_d^p=(32\pm 10){\rm~MeV}\,,
\\
\sigma_u^n=(15\pm 5){\rm~MeV}\,, \qquad \sigma_d^n=(36\pm 10){\rm~MeV}\,.
\end{split}
\end{equation}

For corrections of higher order in chiral counting one would also need
$F_S^{\prime \, q/N}(0)$. These are of the same order, ${\mathcal
  O}(q)$, as the two-nucleon contributions which are not captured in
our expressions.

\subsection{Pseudoscalar current}
In the LO expressions we only need the light meson pole parts of the
pseudoscalar form factor, Eq. \eqref{pseudoscalar:form:factor},
\begin{equation}
\label{eq:F_P:LO}
F_{P}^{q/N}(q^2)=\frac{m_N^2}{m_\pi^2-q^2} a_{P,\pi}^{q/N}+\frac{m_N^2}{m_\eta^2-q^2} a_{P,\eta}^{q/N}+\cdots\,,
\end{equation} 
The residua of the poles are given by
\begin{align}
\frac{a_{P,\pi}^{u/p}}{m_u}&=-\frac{a_{P,\pi}^{d/p}}{m_d}=\frac{B_0}{m_N} g_A\,, \qquad \frac{a_{P,\pi}^{s/p}}{m_s}=0\,,
\\
\frac{a_{P,\eta}^{u/p}}{m_u}&=\frac{a_{P,\eta}^{d/p}}{m_d}=-\frac{1}{2}\frac{a_{P,\eta}^{s/p}}{m_s}=\frac{B_0}{3m_N}\big(\Delta u_p+\Delta d_p-2 \Delta s_p\big)\,,
\end{align}
where the values of the axial-vector elements, $\Delta q$, are given
in~\eqref{eq:App:u-d} and~\eqref{eq:Deltaq:values}. Moreover, $B_0$ is
a ChPT constant related to the quark condensate given, up to
corrections of ${\mathcal O}(m_q)$, by $\langle \bar q q\rangle \simeq
-f^2 B_0$. Using quark condensate from~\cite{McNeile:2012xh} and the
LO relation $f=f_\pi$, with $f_\pi$ the pion decay constant, one has
$B_0=2.666(57) {\rm ~GeV}$, evaluated at the scale $\mu=2\,$GeV.

In practice, $B_0$ never appears by itself, but rather as the product
$B_0 m_q$ which can be expressed in terms of the pion mass and quark
mass ratios, 
\begin{equation}
\begin{split}\label{eq:B0mq:num}
B_0 m_u & = \frac{m_\pi^2}{1+m_d/m_u}=  (6.1\pm0.5) \times 10^{-3} \,\text{GeV}^2\,,\\
B_0 m_d & = \frac{m_\pi^2}{1+m_u/m_d} =  (13.3\pm0.5) \times 10^{-3} \,\text{GeV}^2\,,\\
B_0 m_s & =\frac{m_\pi^2}{2}\frac{m_s}{\bar m} = (268 \pm 3) \times 10^{-3} \,\text{GeV}^2\,.
\end{split}
\end{equation}
The numerical values are obtained using the ratios of quark masses,
$m_u/m_d=0.46\pm0.05$, $m_s/\bar m = 27.5\pm0.3$ (see the quark mass
review in~\cite{Olive:2016xmw}), and the charged-pion mass
$m_\pi$.

At NLO in the chiral expansion, the above expressions for
$a_{P,\pi}^{q/p}$ and $a_{P,\eta}^{q/p}$ get corrections of ${\mathcal
  O}(m_{\pi,\eta}^2/(4\pi f_\pi)^2)$. In addition one needs to keep
the constant term in the $q^2$ expansion of the form factor
\begin{equation}
\label{eq:F_P':NLO}
F_{P}^{q/N}(q^2)=\frac{m_N^2}{m_\pi^2-q^2} a_{P,\pi}^{q/N}+\frac{m_N^2}{m_\eta^2-q^2} a_{P,\eta}^{q/N}+b_{P}^{q/N}+\cdots\,.
\end{equation}
In our numerical analysis we estimate the size of these higher-order
corrections by using the NDA size for
\begin{equation}
b_{P}^{q/N}\approx 1\,, \qquad {\rm where}\quad q=u,d,s\,,
\end{equation}  
while keeping $a_{P,\pi}^{q/p}$, $a_{P,\eta}^{q/p}$ at their LO
values. This treatment of NLO corrections is only approximate, but
suffices for the present precision. Furthermore, it can be improved in
the future.

\subsection{CP-even gluonic current}

The matrix element of the CP-even gluonic
current~\eqref{CPeven:gluonic:form:factor} is parametrized by a single
form factor $F_G^{N}(q^2)$. The LO expressions in chiral counting
require only its value at zero momentum transfer, 
\begin{equation}\label{eq:FG:def}
F_G^{N}(0)=-\frac{2m_G}{27}.
\end{equation}
The nonperturbative coefficient $m_G$ is the gluonic contribution to
the nucleon mass in the isospin limit,
\begin{equation}
m_G \bar u_N u_N=- \frac{9\alpha_s}{8\pi}\langle N |G_{\mu\nu}G^{\mu\nu}|N \rangle\,.
\end{equation}
The trace of the stress-energy tensor,
$\theta_\mu^\mu=-9\alpha_s/(8\pi) G_{\mu \nu}
G^{\mu\nu}+\sum_{u,d,s}m_q\bar q q$, yields the relation
\begin{equation}
\label{eq:mG:value}
m_G=m_N-\sum_q \sigma_q^N=(848\pm14) {\rm ~MeV}\,, 
\end{equation}
where in the last equality we used the values for $\sigma_q$
in~\eqref{eq:values:sigmas} and~\eqref{eq:values:sigmaqN}. While the
isospin violation in the $\sigma_q^N$ values is of ${\mathcal
  O}(10\%)$, this translates to a very small isospin violation in
$m_G$, of less than 1 MeV. The value of $m_G$ in \eqref{eq:mG:value}
thus applies to both $N=p$ and $N=n$.

For the derivative of $F_G$ at zero recoil we use the naive
dimensional analysis estimate
\begin{equation}
\frac{F_G'(0)}{F_G(0)}\approx 1/m_N^2 \approx 1~{\rm GeV}^{-2}\,.
\end{equation}

\subsection{CP-odd gluonic current}
The matrix element of the CP-odd gluonic current
\eqref{CPodd:gluonic:form:factor} is related to the matrix elements of
the axial and pseudoscalar currents through the QCD chiral
anomaly. Namely, a chiral rotation of the quark fields, $q\to
\exp(i\beta\gamma_5 )q$, shifts the QCD theta spurion by $\theta \to
\theta - 2 \Tr\beta$, along with corresponding changes in the
pseudoscalar and axial-vector spurions (see
Ref.~\cite{Bishara:2016hek}).
This implies a relation,
\begin{equation}\label{eq:theta-rel}
\frac{1}{\tilde m}\langle  N'|\frac{\alpha_s}{8 \pi}G_{\mu\nu}^a
\tilde G^{a\mu\nu}|N\rangle=\sum_q \Big(\langle N'| \bar q i\gamma_5
q|N\rangle - \frac{1}{2m_q} \partial_\mu \langle N'|  \bar q
\gamma^\mu \gamma_5 q|N\rangle\Big)\,,  
\end{equation}
valid at leading order in the chiral expansion. To shorten the
notation we defined $1/\tilde m=(1/m_u+1/m_d+1/m_s)$. In terms of form
factors this gives 
\begin{equation}
\frac{1}{\tilde m}F_{\tilde G}^{N}=\sum_q
\Big(\frac{1}{m_q}F_P^{q/N}-\frac{m_N}{m_q}F_A^{q/N}-\frac{q^2}{4 m_N
  m_q} F_{P'}^{q/N}\Big)\,.
\end{equation}
The leading order contributions from $F_P^{q/N}$ cancel in the sum,
  giving
\begin{equation}
\begin{split}\label{FGtilde:LO}
F_{\tilde G}^{N}(q^2)=-\tilde m m_N\Big[&\frac{\Delta u}{m_u}+\frac{\Delta d}{m_d}+\frac{\Delta s}{m_s}+\frac{g_A}{2}\Big(\frac{1}{m_u}-\frac{1}{m_d}\Big)\frac{q^2}{m_\pi^2-q^2}
\\
&+\frac{1}{6}\big(\Delta u+\Delta d-2 \Delta s)\Big(\frac{1}{m_u}+\frac{1}{m_d}-\frac{2}{m_s}\Big)\frac{q^2}{m_\eta^2-q^2}\Big].
\end{split}
\end{equation}
The pion pole contribution would vanish in the exact isospin
limit. However, the isospin breaking effects in the matrix element of
$\tilde G G$ operator are not small~\cite{Gross:1979ur}. This is
unlike most of the other observables, where isospin breaking is
suppressed by the chiral scale, $\propto (m_u-m_d)/(4 \pi
f_\pi)$. Here, the isospin breaking is proportional to
$(m_u-m_d)/(m_u+m_d)\sim {\mathcal O}(1)$ and is thus
large. Similarly, the $\eta$ pole contribution would vanish in the
limit of exact SU(3), but is in fact an ${\mathcal O}(1)$ correction.

The LO expression for $F_{\tilde G}^N$, Eq. \eqref{FGtilde:LO},
contains both the constant term as well as poles of the form $\sim
q^2/(m_\pi^2-q^2)$. At NLO in chiral counting one also has in addition
the ${\mathcal O}(q^2)$ contribution,
\begin{equation}
F_{\tilde G}^{N}(q^2)=
\frac{q^2}{m_\pi^2-q^2} a_{\tilde G,\pi}^{N}+\frac{q^2}{m_\eta^2-q^2} a_{\tilde G,\eta}^{N}+b_{\tilde G}^{N}
+c_{\tilde G}^{N} q^2+\cdots.
\end{equation}
At NLO the $a_{\tilde G,\pi}^{N}, a_{\tilde G,\eta}^{N}, b_{\tilde
  G}^{N}$ coefficients differ from their LO values in
\eqref{FGtilde:LO} by relative correction of the size ${\mathcal
  O}\big(m_{\pi,\eta}^2/(4\pi f_\pi)^2\big)$, while the NDA estimate
for the NLO coefficient is $c_{\tilde G}^N\approx 1$.

\subsection{Tensor current}
The matrix element of the tensor current~\eqref{tensor:form:factor} is
described by three form factors, $F_{T,0}^{q/N}(q^2)$,
$F_{T,1}^{q/N}(q^2)$, $F_{T,2}^{q/N}(q^2)$. These are related to the
generalized tensor form factors through (see, e.g.,
\cite{Gockeler:2005cj,Diehl:2001pm}) 
\begin{align}
F_{T,0}^{q/N}(q^2)&=m_q A_{T,10}^{q/N}(q^2)\,, \label{eq:FT0:def}
\\
 F_{T,1}^{q/N}(q^2)&= - m_q B_{T,10}^{q/N} (q^2)\,, \label{eq:FT1:def}
 \\
F_{T,2}^{q/N}(q^2)&=\frac{m_q}{2} \tilde A_{T,10}^{q/N}(q^2)\,.
\end{align}

In the LO expressions for DM scattering only $F_{T,0}^{q/N}(0)$ and
$F_{T,1}^{q/N}(0)$ appear. The value of $F_{T,0}^{q/N}(0)$ is quite
well determined. A common notation is $A_{T,10}^{q/p}(0)=g_T^q$ (with
$A_{T,10}^{u(d)/p}=A_{T,10}^{d(u)/n}$ and
$A_{T,10}^{s/p}=A_{T,10}^{s/n}$ in the isospin limit), so that
\begin{equation}\label{eq:FT0:tqN}
F_{T,0}^{q/p}(0)=m_q  g_{T}^{q}.
\end{equation}
The tensor charges are related to the transversity structure functions
$\delta q_N(x,\mu)$ by $g_{T}^q(\mu)=\int_{-1}^{1}dx \delta
q_N(x,\mu)$. These structure functions can, in principle, be measured
in deep inelastic scattering, but this determination is not very
precise. Recent lattice calculations include both connected and
disconnected contributions and give, in the $\overline{\rm MS}$ scheme
at $\mu=2$~GeV \cite{Alexandrou:2017qyt, 
  Bhattacharya:2016zcn}, 
\begin{equation}\label{eq:tqN:val}
g_T^u=0.794\pm0.015\,, \qquad g_T^d=-0.204\pm0.008\,, \qquad g_T^s=(3.2\pm8.6)\cdot 10^{-4}\,.
\end{equation}
This agrees well with previous, less precise, determinations
\cite{Gockeler:2005cj, Pleiter:2011gw, Bali:2014nma,
  Bhattacharya:2015wna, Lin:2008uz, Green:2012ej, Alexandrou:2013wka,
  Abdel-Rehim:2015owa}.  It is interesting to compare
\eqref{eq:tqN:val} with the results from the constituent quark model
\cite{Pasquini:2005dk}, $g_T^u=0.97$, $g_T^d=-0.24$, as we will have
to use this model below. In the nonrelativistic quark model, on the
other hand, using just $SU(6)$ spin-flavor symmetry, one gets
$g_T^u=4/3$, $g_T^d=-1/3$, see, e.g., \cite{Gao:2017ade}.

The zero recoil values of the other two form factors,
$F_{T,1}^{q/N}(0)$ and $F_{T,2}^{q/N}(0)$, are less well
determined. The constituent quark model of~\cite{Pasquini:2005dk}
gives 
\begin{align}
B_{T,10}^{u/p}(0)\approx 3.0\,, &\qquad \tilde A_{T,10}^{u/p}\approx -0.50\,,\label{eq:BT10u}
\\
B_{T,10}^{d/p}(0)\approx 0.24\,, &\qquad\tilde A_{T,10}^{d/p}\approx 0.46\,.\label{eq:BT10d}
\end{align}
The form factors for the neutron are obtained through the replacements
$u\leftrightarrow d$, $p \to n$. We assign a $50\%$ error to the above
estimates, taking as a guide twice the difference between the
determination of $g_T^q$ in this model and in lattice QCD
\eqref{eq:tqN:val}.  For the $s$ quark we use the very rough estimates
\begin{equation}\label{eq:BtildeA:s}
-0.2 \lesssim B_{T,10}^{s/p}(0)\,, \tilde A_{T,10}^{s/p}(0)\lesssim 0.2\,.
\end{equation}

The linear combination
\begin{equation}
  \kappa_T^q=2 \tilde A_{T,10}^{q/p}(0)+B_{T,10}^{q/p}(0)
\end{equation}
is in fact much better known than $\tilde A_{T,10}(0)$ and
$B_{T,10}(0)$ separately. The tensor magnetic moments, $\kappa_T^q$,
for the $u$ and $d$ quarks were determined using lattice QCD to be, at
$\mu=2$ GeV~\cite{Gockeler:2006zu},
\begin{equation}\label{eq:kappaTq}
  \kappa_T^u\approx 3.0\,, \qquad \kappa_T^d\approx 1.9\,
\end{equation}
(no uncertainty is given in this reference). In the constituent quark
model of~\cite{Pasquini:2005dk} one gets $\kappa_T^u\approx 2.0$,
$\kappa_T^d\approx 1.2$, which agrees with \eqref{eq:kappaTq} within
the assigned 50\% uncertainty (larger values $\kappa_T^u=3.60,
\kappa_T^d=2.36$ are obtained with a simple harmonic oscillator wave
function \cite{Schmidt:1997vm,Pasquini:2005dk}). For the strange quark
one obtains from the SU(3) chiral quark-soliton model
\cite{Ledwig:2010zq}
\begin{equation}
  -0.2 \lesssim\kappa_T^s\lesssim 0.2,
\end{equation}
motivating the ranges in \eqref{eq:BtildeA:s} (in~\cite{Ledwig:2011qw}
a much smaller value $\kappa_T^s\approx 0.01$ was found.

In Refs.~\cite{Gockeler:2005cj, Alexandrou:2013wka, Zanotti:2017bte},
lattice QCD results for the $q^2$ dependence of $F_{T,0}^{q/N}$ for
$u$ and $d$ quarks were presented.
Averaging over them gives
\begin{equation}\label{eq:FT0der}
\frac{F_{T,0}^{u/p}{}'(0)}{F_{T,0}^{u/p}(0)}\approx (0.8\pm 0.3)~{\rm
  GeV}^{-2}\,, \qquad
\frac{F_{T,0}^{d/p}{}'(0)}{F_{T,0}^{d/p}(0)}\approx(0.7\pm0.2)~{\rm
  GeV}^{-2}\,, 
\end{equation}
where the errors reflect the differences between the three determinations.
For the $s$-quark form factor one can use the NDA estimate, 
${F_{T,0}^{s/p}{}'(0)}/{F_{T,0}^{s/p}(0)}\approx1~{\rm
  GeV}^{-2}$, consistent with the above.

For the other two form factors an estimate of the derivative at zero
recoil can be made using the results from the constituent quark model
of \cite{Pasquini:2005dk}, giving 
\begin{align}
\frac{F_{T,1}^{u/p}{}'(0)}{F_{T,1}^{u/p}(0)}&\approx 1.0 ~{\rm GeV}^{-2}\,, \qquad \frac{F_{T,1}^{d/p}{}'(0)}{F_{T,1}^{d/p}(0)}\approx -0.1~{\rm GeV}^{-2}\,,
\\
\frac{F_{T,2}^{u/p}{}'(0)}{F_{T,2}^{u/p}(0)}&\approx 1.2 ~{\rm GeV}^{-2}\,, \qquad \frac{F_{T,2}^{d/p}{}'(0)}{F_{T,2}^{d/p}(0)}\approx 1.0~{\rm GeV}^{-2}\,.
\end{align}
These estimates most probably have large errors, since within this
model one gets $F_{T,0}^{u/p}{}'(0)/F_{T,0}^{u/p}(0)\approx0.22~{\rm
  GeV}^{-2}$, $F_{T,0}^{d/p}{}'(0)/F_{T,0}^{d/p}(0)\approx0.24~{\rm
  GeV}^{-2}$, 
about a factor of three smaller than lattice QCD determination in
\eqref{eq:FT0der}. For the strange quark form factor we vary the
derivative at zero recoil in the range
\begin{equation}
-2~{\rm GeV}^{-2}\lesssim F_{T,1}^{s/p}{}'(0)\,, F_{T,2}^{s/p}{}'(0) \lesssim 2~{\rm GeV}^{-2}\,,
\end{equation}
motivated by the slope $d\kappa_T^s/dq^2\approx -2.2~{\rm GeV}^{-2}$
that one can deduce from the results in \cite{Ledwig:2011qw}.

\section{Nonrelativistic expansion of currents for fermions}
\label{app:NR}
In this appendix we give the nonrelativistic expansion of the DM and nucleon
currents. We first focus on fermionic DM and then translate the results to
nonrelativistic nucleons.
In order to get rid of the time derivative, $v\cdot \partial$, in the
higher-order terms in the Heavy Dark Matter Effective Theory (HDMET)
Lagrangian, the tree level relation
\begin{equation}
\chi=e^{-i m_\chi v \cdot x} \Big(1 +\frac{i \slashed
  \partial_\perp}{i v\cdot \partial+2 m_\chi-i \epsilon}\Big)
\chi_v\,,\label{eq:chi:rel}
\end{equation}
is supplemented with a field redefinition\footnote{In order for the
  scattering rates to be independent of this arbitrary field
  redefinition, contributions to the scattering amplitude from the
  time-ordered product of the Lagrangians~\eqref{eq:LNR}
  and~\eqref{eq:L:NRQED} have to be included~\cite{Falk:1993dh}. An
  explicit calculation shows that, with our
  choice~\eqref{eq:field:redef}, these additional contributions vanish
  to ${\mathcal O}(p^2)$.}~\cite{Manohar:1997qy}
\begin{equation}
\label{eq:field:redef}
\chi_v\to \Big(1-\frac{\partial_\perp^2}{8 m_\chi^2}+\frac{\partial_\perp^2 (i v\cdot \partial)}{16 m_\chi^3}+\cdots \Big) \chi_v\,,
\end{equation}
where $\partial_\perp^\mu=\partial^\mu-v\cdot \partial\, v^\mu$. In
this way one obtains the conventional ``NRQED'' Lagrangian,
\begin{equation}\label{eq:L:NRQED}
{\cal L}_{\rm NRQED}=\chi_v^\dagger \Big( i v\cdot \partial+\frac{ (i \partial_\perp)^2}{2 m_\chi}+ \frac{ (i \partial_\perp)^4}{8 m_\chi^3}+\cdots \Big)\chi_v,
\end{equation}
also beyond ${\mathcal O}(p^2)$ order. 

Using~\eqref{eq:chi:rel} together with~\eqref{eq:field:redef} and
applying the equation of motion derived from Eq.~\eqref{eq:L:NRQED} we
obtain for the DM currents 
\begin{align}
\label{eq:HDMETlimit:scalar}
\bar \chi \chi&\to \bar \chi_v \chi_v+\frac{i}{2 m_\chi^2}\epsilon_{\alpha\beta\mu\nu}v^\alpha \big( \bar \chi_v S_{\chi}^\beta  \lpartial_{\perp}^{\mu} \rpartial_{\perp}^{\nu}\chi_v\big) -\frac{1}{8m_\chi^2}\bar \chi_v \lrpartial_\perp^2\chi_v+{\mathcal O}(p^3)\,,
\\
\label{eq:HDMETlimit:pscalar}
\begin{split}
\bar \chi i \gamma_5 \chi &\to \frac{1}{ m_\chi}\partial_\mu \big(\bar
\chi_v S_\chi^\mu \chi_v \big)  
\\
&\quad -\frac{1}{4m_\chi^3}\partial_\perp^\mu\bar \chi_v
S_{\chi,\mu}\big(\lpartial_\perp^2+\rpartial_\perp^2\big)\chi_v 
 +\frac{1}{8m_\chi^3} \chi_v S_{\chi} \ncdot \lrpartial_\perp
 \big(\lpartial_\perp^2-\rpartial_\perp^2\big)\chi_v  
 +{\mathcal O}(p^4)\,,
\end{split}
\\
\label{eq:vecDM:expand}
\begin{split}
\bar \chi \gamma^\mu \chi &
\to  v^\mu \bar \chi_v \chi_v +\frac{1}{2 m_\chi}\bar \chi_v i \lrpartial_{\perp}^{\mu} \chi_v
+\frac{1}{2 m_\chi} \partial_\nu\big(\bar \chi_v \sigma_\perp^{\mu\nu} \chi_v\big)
\\
&\quad+\frac{i}{4 m_\chi^2} v^\mu \bar \chi_v \lpartial_{\rho}\sigma_\perp^{\rho\nu} \rpartial_{\nu}\chi_v -\frac{v^\mu}{8 m_\chi^2}\partial_\perp^2 \bar \chi_v\chi_v 
\\
&\quad +\frac{1}{16 m_\chi^3}\Big(i \partial^\mu \big(\bar \chi_v \big( 
\lpartial_\perp^2 - \rpartial_\perp^2 \big) \chi_v \big) - 2 \bar \chi_v \big( 
\lpartial_\perp^2 + \rpartial_\perp^2 \big) i \lrpartial^\mu \chi_v
\\ & \qquad \qquad \quad - \bar \chi_v \big(
\rpartial_\perp^2-\lpartial_\perp^2\big)
\sigma_\perp^{\mu\nu}\lrpartial_\perp^\nu\chi_v - 2 \partial^\nu \big(
\bar \chi_v \big(\rpartial_\perp^2+\lpartial_\perp^2\big)
\sigma_\perp^{\mu\nu}\chi_v \big) \Big)+{\mathcal O}(p^4)\,, 
\end{split}
\\
\label{eq:axialDM:expand}
\begin{split}
\bar \chi \gamma^\mu \gamma_5 \chi & 
\to  2 \bar \chi_v S_\chi^\mu \chi_v -\frac{i}{m_\chi} v^\mu \bar \chi_v S_\chi\cdot \lrpartial \chi_v
\\
&\quad-\frac{1}{4m_\chi^2}\bar \chi_v \lrpartial_\perp^2S_\chi^\mu \chi_v-\frac{1}{2 m_\chi^2}\bar \chi_v \big(\lpartial_\perp^\mu S\ncdot\partial_\perp+\lpartial_\perp\ncdot S \partial_\perp^\mu\big)\chi_v
\\
&\quad+\frac{i}{4 m_\chi^2}\varepsilon^{\mu\nu\alpha\beta}v_\nu\bar \chi_v \lpartial_{\perp\alpha}\partial_{\perp\beta}\chi_v-\frac{i}{8 m_\chi^3}v^\mu \partial_\nu \bar \chi_v \big(\lpartial_\perp^2-\rpartial_\perp^2\big) S_\chi^\nu \chi_v
\\
&\quad+\frac{i}{4 m_\chi^3}v^\mu  \bar \chi_v \big(\lpartial_\perp^2+\rpartial_\perp^2\big) \lrpartial \ncdot S_\chi \chi_v+\mathcal{O}(p^4)\,,
\end{split}
\\
\label{eq:tensorDM:expand}
\begin{split}
\bar \chi \sigma^{\mu\nu} \chi& \to  \bar \chi_v \sigma_\perp^{\mu\nu}
\chi_v + \frac{1}{2m_\chi} \Big( \bar\chi_v i v_{\phantom{\perp}}^{[\mu}
  \sigma_\perp^{\nu]\rho} \lrpartial_{\rho} \chi_v - v^{[\mu}
  \partial^{\nu]} \bar\chi_v \chi_v\Big) 
  \\
  &\quad+\frac{1}{4m_\chi^2}\bar \chi_v  \slpartial_\perp
  \sigma_\perp^{\mu\nu}\srpartial_\perp \chi_v
  -\frac{1}{8m_\chi^2}\bar\chi_v
  (\,\lpartial_\perp^2+\rpartial_\perp^2) \sigma_\perp^{\mu\nu} \chi_v+ {\mathcal O}(p^3)\,,
  \end{split}
  \\
  \label{eq:axialtensorDM:expand}
  \begin{split}
  \bar
\chi \sigma^{\mu\nu} i\gamma_5 \chi&\to 
 2\bar
\chi_v S_\chi^{[\mu}v^{\nu]} \chi_v+\frac{i}{m_\chi}\bar \chi_v S^{[\mu}\lrpartial_\perp^{\nu]}\chi_v+\frac{1}{2 m_\chi}\epsilon^{\mu\nu\alpha\beta}v_\alpha \partial_{\perp\beta}\bar\chi_v \chi_v
\\
&\quad+\frac{1}{4 m_\chi^2}  \partial_\perp^2 \bar \chi_v v^{[\mu}S_\chi^{\nu]}\chi_v +\frac{1}{2 m_\chi^2}\bar \chi_v\lpartial_\perp^{[\mu}v^{\nu]}S_\chi\cdot\rpartial_\perp \chi_v+
\frac{1}{2 m_\chi^2}\bar \chi_v\lpartial_\perp\ncdot S_\chi \rpartial_\perp^{[\mu} v^{\nu]} \chi_v 
\\
&\quad+\frac{i}{4m_\chi^2}v^{[\mu}\epsilon^{\nu] \alpha\beta\gamma}\bar \chi_v \lpartial_{\perp\alpha}\rpartial_{\perp\beta}v_\gamma \chi_v +{\mathcal O}(p^3)\,, 
\end{split}
\end{align}
where $\sigma_\perp^{\mu\nu}=i [\gamma_\perp^\mu,
  \gamma_\perp^\nu]/2$, $\bar \chi_v \lrpartial^\mu \chi_v=\bar \chi_v
(\partial^\mu \chi_v)- (\partial^\mu\bar \chi_v) \chi_v$, and
$S^\mu=\gamma_\perp^\mu \gamma_5/2 $ is the spin operator. The square
brackets in the last line denote antisymmetrization in the enclosed
indices, while the ellipses denote higher orders in $1/m_\chi$. We
also used the relation
\begin{equation}\label{eq:sigma-to-epsilon-S}
\bar \chi_v \sigma_\perp^{\mu\nu}\chi_v=
-2 \epsilon^{\mu\nu\alpha\beta}v_{\alpha}\big(\bar \chi_v S_{\chi,\beta}\chi_v\big)\,,
\end{equation}
where $\epsilon^{\mu\nu\alpha\beta}$ is the totally antisymmetric
Levi-Civita tensor, with $\epsilon^{0123}=1$, and
\begin{equation}
\bar \chi_v S^\mu \ncdot S^\nu \chi_v = -\tfrac{i}{2}
\epsilon^{\mu\nu\alpha\beta} \bar \chi_v v_\alpha S_\beta \chi_v -
\tfrac{1}{4} \bar \chi_v g_\perp^{\mu\nu} \chi_v\,.
\end{equation}
The same expressions apply also for nucleon currents, with the obvious
replacement $\chi \to N$. In terms of the momenta instead of
derivatives the expansions are
\begin{align}
\label{eq:HDMETlimit:scalar:momenta}
\bar \chi \chi&\to \bar \chi_v \chi_v\Big(1+\frac{p_{12}^2}{8 m_\chi^2}\Big)+
\frac{i}{2 m_\chi^2}\epsilon_{\alpha\mu\nu\beta}v^\alpha p_{2}^{\mu} p_{1}^{\nu}\big( \bar \chi_v S_{\chi}^\beta  \chi_v\big) +{\mathcal O}(p^3)\,,
\\
\label{eq:HDMETlimit:pscalar:momenta}
\begin{split}
 \bar \chi i \gamma_5
\chi &\to  \frac{-i}{ m_\chi}\big(\bar \chi_v q \ncdot S_\chi \chi_v \big)\Big(1+\frac{p_{1}^2+p_{2}^2}{4 m_\chi^2}\Big) 
\\
&\qquad + \frac{i}{8 m_\chi^3}  \big(p_2^2-p_{1}^2 \big) \bar
  \chi_v \big(S_\chi\ncdot p_{12} \big) \chi_v + {\mathcal O}(p^4)\,,
\end{split}
\\
\label{eq:vecDM:expand:momenta}
\begin{split}
\bar \chi \gamma^\mu \chi &
\to   \bar \chi_v \chi_v \Big( v^\mu +\frac{p_{12,\perp}^\mu}{2m_\chi}
+v^\mu\frac{q_\perp^2}{8 m_\chi^2}\Big)+\frac{i}{m_\chi}\epsilon^{\alpha\mu\nu\beta}v_\alpha q_{\nu} \big(\bar \chi_v S_{\chi,\beta} \chi_v\big)
\\
&\quad-\frac{i}{2 m_\chi^2} v^\mu \epsilon^{\alpha\rho\nu\beta} v_\alpha p_{2\rho}p_{1\nu} \big(\bar \chi_v S_{\chi,\beta} \chi_v\big)
\\
&\quad +\frac{1}{16 m_\chi^3} \Big[ q^\mu
\big(p_{1\perp}^2-p_{2\perp}^2 \big) + 2 p_{12}^\mu
\big(p_{1\perp}^2+p_{2\perp}^2 \big) \Big] \bar \chi_v
\chi_v 
\\
&\quad + \frac{i}{8 m_\chi^3} \Big[ p_{12,\nu} \big(p_{1\perp}^2 -
  p_{2\perp}^2 \big) + 2 q_\nu \big(p_{1\perp}^2 + p_{2\perp}^2 \big)
  \Big] \epsilon^{\mu\nu\alpha\beta} v_\alpha \bar\chi_v S_{\chi,
  \beta} \chi_v + {\mathcal O}(p^4)\,,   
\end{split}
\\
\label{eq:axialDM:expand:momenta}
\begin{split}
\bar \chi \gamma^\mu \gamma_5 \chi & 
\to  2 \bar \chi_v S_\chi^\mu \chi_v \Big(1+\frac{p_{12\perp}^2}{8 m_\chi^2}\Big)-\frac{1}{m_\chi} v^\mu \bar \chi_v S_\chi\ncdot p_{12} \chi_v
\\
&\quad -\frac{1}{4 m_\chi^2}\bar \chi_v \big(p_{12\perp}^\mu
  S_\chi\ncdot p_{12} - q_{\perp}^\mu S_\chi \ncdot q \big)
  \chi_v -\frac{i}{4 m_\chi^2}\varepsilon^{\nu\mu\alpha\beta}v_\nu
p_{2\alpha}p_{1\beta}\bar \chi_v \chi_v 
\\
&\quad -\frac{v^\mu}{8m_\chi^3} \bar \chi_v \Big[
   \big( p_{1\perp}^2-p_{2\perp}^2\big) q\ncdot S_\chi +2 \big(p_{1\perp}^2 + 
    p_{2\perp}^2\big) p_{12} \ncdot S_\chi \Big] \chi_v + 
  \mathcal{O}(p^4)\,, 
\end{split}
\\
\label{eq:tensorDM:expand:momenta}
\begin{split}
\bar \chi \sigma^{\mu\nu} \chi& \to  -2
\varepsilon^{\mu\nu\alpha\beta} v_\alpha \big(\bar \chi_v
S_{\chi,\beta} \chi_v \big) \Big(1+\frac{p_{12}^2}{8 m_\chi^2}\Big) +
\frac{1}{m_\chi} v^{[\mu}\varepsilon^{\nu]\delta\alpha\beta} v_\delta
p_{12,_\alpha} \bar \chi_v S_{\chi,\beta} \chi_v  
\\
&\qquad+\frac{i}{2 m_\chi} v^{[\mu}q^{\nu]}\bar \chi_v\chi_v +
\frac{i}{4m_\chi^2} p_1^{[\mu}p_2^{\nu]}\bar \chi_v   \chi_v 
\\
&\quad +\frac{1}{2 m_\chi^2} \varepsilon^{\mu\nu\alpha\beta}v_\alpha \bar\chi_v
  \big(p_{1\beta} S_\chi\ncdot p_2+p_{2\beta} S_\chi\ncdot p_1\big) \chi_v+ {\mathcal O}(p^3)\,,
\end{split}
\\  \label{eq:axialtensorDM:expand:momenta}
\begin{split}
  \bar \chi \sigma^{\mu\nu} i\gamma_5 \chi&\to 
 2\bar \chi_v S_\chi^{[\mu}v^{\nu]} \chi_v\Big(1+\frac{q_\perp^2}{8 m_\chi^2}\Big)+\frac{1}{m_\chi}\bar \chi_v S_\chi^{[\mu}p_{12,\perp}^{\nu]}\chi_v-\frac{i}{2 m_\chi}\epsilon^{\mu\nu\alpha\beta}v_\alpha q_\beta\bar\chi_v \chi_v
\\
&\quad +\frac{1}{2 m_\chi^2}\bar \chi_v\big(p_{1}^{[\mu}v^{\nu]}S_\chi\ncdot p_2+p_{2}^{[\mu}v^{\nu]}S_\chi\ncdot p_1\big) \chi_v
\\
&\quad -\frac{i}{4m_\chi^2}v^{[\mu}\epsilon^{\nu] \delta \alpha\beta} v_\delta p_{1\alpha}p_{2\beta}\bar \chi_v  \chi_v+{\mathcal O}(p^3)\,,
\end{split}
\end{align}
where we used the shorthand notation $p_{12}^\mu=p_1^\mu+p_2^\mu$. The
corresponding expansion of the nucleon currents is obtained through
the replacements $\chi\to N$, $p_{1,2}^\mu\to k_{1,2}^\mu$, $q^\mu\to
-q^\mu$.

\section{NLO expressions for fermionic DM}\label{sec:NLOva}

At NLO in the chiral expansion for the hadronization of the
relativistic operators,
Eqs. \eqref{eq:dim6EW:Q1Q2:light}-\eqref{eq:dim7EW:Q9Q10:light}, one
encounters terms that are not Galilean invariant, since they depend on
the average nucleon velocity,
\begin{equation}
\vec v_a=\frac{1}{2 m_N}\big(\vec k_1+\vec k_2\big).
\end{equation} 
These terms signal that the underlying theory is, in fact, Lorentz
rather than Galilean invariant.

In addition to the nonrelativistic operators
\eqref{eq:O1pO2p}-\eqref{eq:O15p} there are three new operators of
${\mathcal O}(q)$,
\begin{align}
\label{eq:O1aO1a1}
{\mathcal O}_{1a}^{N(1)}&= \mathbb{1}_\chi \, \big( \vec v_a \cdot \vec S_N \big) \,,
&{\mathcal O}_{2a}^{N(1)}&= \big( \vec v_a \cdot \vec S_\chi \big) \mathbb{1}_N \,,
\\
\label{eq:O1a3}
{\mathcal O}_{3a}^{N(1)}&= \vec v_a \cdot \Big(\vec S_\chi \times \vec S_N \Big) \,,
&&
\end{align}
four new operators of ${\mathcal O}(q^2)$,
\begin{align}
\label{eq:O1aO2a2}
{\mathcal O}_{1a}^{N(2)}&= \Big( \frac{i\vec q}{m_N} \ncdot \vec S_\chi \Big) \, \big( \vec v_a \cdot \vec S_N \big) \,,
&{\mathcal O}_{2a}^{N(2)}&= \big( \vec v_a \cdot \vec S_\chi \big) \, \Big( \frac{i\vec q}{m_N} \ncdot \vec S_N \Big) \,,
\\
\label{eq:O3aO4a2}
{\mathcal O}_{3a}^{N(2)}&= \big( \vec v_a \cdot \vec S_\chi \big) \, \big( \vec v_a \cdot \vec S_N \big)\,,
&{\mathcal O}_{4a}^{N(2)}&= \Big( \frac{i\vec q}{m_N} \ncdot \vec S_\chi \Big) \, \Big( \frac{i\vec q}{m_N} \ncdot \vec S_N \Big) \,,
\end{align}
and three of ${\mathcal O}(q^3)$, 
\begin{align}
\label{eq:O1aO2a3}
{\mathcal O}_{1a}^{N(3)}&= \big( \vec v_a \ncdot \vec S_\chi \big) \, \vec v_a \cdot \Big(\vec v_\perp \times \vec S_N \Big) \,,
&{\mathcal O}_{2a}^{N(3)}&= \vec v_a \cdot \Big(\vec v_\perp \times \vec S_\chi \Big) \, \big( \vec v_a \ncdot \vec S_N \big) \,,
\\
\label{eq:O3a3}
{\mathcal O}_{3a}^{N(3)}&= \Big( \frac{i\vec q}{m_N} \ncdot \vec S_N \Big) \Big(\,\frac{i\vec q}{m_N} \ncdot \big( \vec v_a \times \vec S_\chi \big) \Big)\,.
&&
\end{align}

Next we give the expressions for the nonrelativistic reduction of the
operators~\eqref{eq:dim6EW:Q1Q2:light}-\eqref{eq:dim7EW:Q9Q10:light}
to subleading order in $q^2$. For each of the operators we stop at the
order at which one expects the contributions from the two-nucleon
currents. We explicitly include a factor 
\begin{equation}\label{eq:norm}
\sqrt{\frac{E_{p_1}E_{p_2}E_{k_1}E_{k_2}}{m_\chi^2 m_N^2}} = 1 +
\frac{\vec q^{\,\,2}}{8} \Big( \frac{1}{m_\chi^2} + \frac{1}{m_N^2}
\Big) + \frac{1}{2} \vec v_\perp^{\,\,2} + \vec v_a^{\,\,2} +
    {\mathcal O}(\vec q^{\,\,4}),
\end{equation}
in order to convert from the usual relativistic normalization of
states, $\langle \chi(p') | \chi(p) \rangle = 2 E_{\vec p} (2\pi)^3
\delta^3 (\vec{p}' - \vec{p})$, where $E_{\vec p} = \sqrt{{\vec p}^2 +
  m_\chi^2}$, to the normalization used in~\cite{Anand:2013yka}. The
hadronization of the dimension-six interaction operators, including
the subleading orders for single-nucleon currents, are then given by,
\begin{align}
\begin{split}
\label{eq:Q1q6:NLO}
\Q_{1,q}^{(6)}\to& F_1^{q/N}\op_1^N+\Big\{F_1^{q/N} \frac{\vec
  v_\perp^{\,\,2}}{2} \op_1^N -F_2^{q/N} \frac{\vec q^{\,\,2}}{4 m_N^2}\op_1^N
-\big(F_1^{q/N}+F_2^{q/N}\big)\frac{\vec q^{\,\,2}}{m_\chi m_N}\op_4^N
\\
& +\big( F_1^{q/N}+F_2^{q/N}\big)\op_3^N
+\frac{m_N}{2 m_\chi} F_1^{q/N}\op_5^N+\frac{m_N}{ m_\chi}\Big(F_1^{q/N}+ F_2^{q/N}\Big)\op_6^N +{\mathcal O}(q^2)\Big\}\,,
\end{split}
\\
\begin{split}
\label{eq:Q2q6:NLO}
\Q_{2,q}^{(6)}\to& 2 F_1^{q/N}\op_8^N+ 2\big(F_1^{q/N}+F_2^{q/N}\big) \op_9^N+{\mathcal O}(q^2)\,,
\end{split}
\\
\begin{split}
\Q_{3,q}^{(6)}\to& -2 F_A^{q/N} \Big(\op_7^N- \frac{m_N}{m_\chi} \op_9^N\Big)
-\Big\{ F_A^{q/N} \Big(\op_7^N- \frac{m_N}{m_\chi} \op_9^N\Big)
\frac{\vec q^{\,\,2}}{4 m_N^2}
\\
&- F_A^{q/N} \Big( \big( \vec v_a \ncdot \vec v_\perp \big)
\op_{1a}^{N(1)} + \frac{i\vec q \ncdot \vec v_a}{m_\chi} \op_{3a}^{N(1)}
  \Big) 
  + \frac{1}{2} F_{P'} \frac{i\vec q \ncdot \vec v_a}{m_N}
 \big( \vec v_a \ncdot \vec v_\perp \big)\op_{10}^{N} + {\mathcal O}(q^4)\Big\}\,,
\end{split}
\\
\begin{split}
\Q_{4,q}^{(6)}\to& -4 F_A^{q/N} \op_4^N+F_{P'}^{q/N}\op_6^N
-\Big\{\frac{\vec q^{\,\,2}}{2} F_A^{q/N} \op_4^N\Big(\frac{1}{m_\chi^2}
+ \frac{1}{m_N^2} \Big)
\\
&-\frac{1}{2}F_A^{q/N}\Big(1+\frac{m_N^2}{m_\chi^2}\Big)\op_6^N
-\frac{m_N}{2 m_\chi}F_A^{q/N}\op_3^N+2 F_A^{q/N}\op_{2b}^N
\\
&-\frac{1}{2}F_{P'} \, \frac{i \vec q \ncdot \vec v_a}{m_N}\,
\Big(\op_{1a}^{N(2)} + \op_{2a}^{N(2)}\Big)+{\mathcal O}(q^3)\Big\}\,.
\end{split}
\end{align}
The terms in the curly brackets arise for the first time at subleading
order, i.e., at ${\mathcal O}(q^{\nu_{\rm LO}+2})$. The form factors
in these expressions are evaluated at $q^2=0$, i.e., $F_i\to
F_i(0)$. In the LO terms, on the other hand, one should expand the
form factors to ${\mathcal O}(q^2)$, i.e., in the expressions outside
curly brackets, $F_{i}\to F_{i}(0)+F_{i}'(0)q^2$.

Note that the hadronization of $\Q_{1,q}^{(6)}$ is expected to receive
contributions from two-nucleon currents at ${\mathcal O}(q^2)$, i.e.,
at the same order as the displayed corrections from the single-nucleon
current.  In the hadronization of $\Q_{2,q}^{(6)}$ we do not show the
subleading corrections from expanding the single-nucleon currents. In
this case the two-nucleon currents enter at ${\mathcal O}(q^2)$, while
the higher-order corrections from single-nucleon currents start only
at ${\mathcal O}(q^3)$. Note also that, at $\mathcal{O}(p^4)$, the
hadronization of $\Q_{4,q}^{(6)}$ receives a contribution that is
coherently enhanced, but suppressed by a numerical factor $\sim
1/(16m_N m_\chi)$.

The hadronizations of the dimension-seven operators are given by 
\begin{align} 
\begin{split}\label{eq:Q17:had}
\Q_{1}^{(7)}\to& F_G^{N}\op_1^N+\Big\{F_G^{N}\frac{\vec q^{\,\,2}}{8}\Big(\frac{1}{m_\chi^2}
+ \frac{1}{m_N^2} \Big)\op_1^N-\frac{m_N}{2m_\chi}F_G^{N} \op_5^N+{\mathcal O}(q^3)\Big\}\,,
\end{split}
\\
\begin{split}\label{eq:Q27:had}
\Q_{2}^{(7)}\to& -\frac{m_N}{m_\chi}F_G^{N}\op_{11}^N
-\Big\{\frac{\vec q^{\,\,2}}{8 m_N m_\chi} F_G^{N}\op_{11}^N 
+ \frac{i \vec q \ncdot \vec v_a}{m_\chi}\,F_G^{N} \op_{2a}^{N(1)} + {\mathcal O}(q^4)\Big\}\,,
\end{split}
\\
\begin{split}\label{eq:Q37:had}
\Q_{3}^{(7)}\to& F_{\tilde G}^{N}\op_{10}^N +\Big\{\frac{\vec
  q^{\,\,2}}{8 m_\chi^2} F_{\tilde G}^{N}\op_{10}^N
+\frac{m_N}{2 m_\chi} F_{\tilde G}^{N}\Big(
\op_{15}^N+\frac{\vec q^{\,\,2}}{m_N^2}\op_{12}^N\Big) 
\\
&+ \frac{i \vec q \ncdot \vec
  v_a}{2m_N}\, F_{\tilde G}^{N} \op_{1a}^{N(1)} + {\mathcal O}(q^4)\Big\}\,,
\end{split}
\\
\begin{split}\label{eq:Q47:had}
\Q_{4}^{(7)}\to& \frac{m_N}{m_\chi}F_{\tilde G}^{N}\op_{6}^N +\Big\{
 \frac{i \vec q \ncdot \vec v_a}{2m_\chi}\, F_{\tilde G}^{N} \Big(
\op_{1a}^{N(2)} + \op_{2a}^{N(2)} \Big) + {\mathcal O}(q^5)\Big\}\,,
\end{split}
\\
\begin{split}\label{eq:Q5q7:had}
\Q_{5,q}^{(7)}\to& F_S^{q/N}\op_1^N+{\mathcal O}(q)\,,
\end{split}
\\
\begin{split}\label{eq:Q6q7:had}
\Q_{6,q}^{(7)}\to& -\frac{m_N}{m_\chi}F_S^{q/N}\op_{11}^N +{\mathcal O}(q^2)\,,
\end{split}
\\
\begin{split}\label{eq:Q7q7:had}
\Q_{7,q}^{(7)}\to& F_{P}^{q/N}\op_{10}^N  +\Big\{\frac{\vec
  q^{\,\,2}}{8 m_\chi^2} F_{P}^{q/N}\op_{10}^N
+\frac{m_N}{2 m_\chi}  F_{P}^{q/N} \Big( \op_{15}^N+\frac{\vec
  q^{\,\,2}}{m_N^2}\op_{12}^N\Big) 
\\
&+ \frac{i \vec q \ncdot \vec
  v_a}{2m_N}\, F_{P}^{N} \op_{1a}^{N(1)} + {\mathcal O}(q^4)\Big\}\,,
\end{split}
\\
\begin{split}\label{eq:Q8q7:had}
\Q_{8,q}^{(7)}\to& \frac{m_N}{m_\chi}F_{P}^{q/N}\op_{6}^N + \Big\{
 \frac{i \vec q \ncdot \vec v_a}{2m_\chi}\, F_{P}^{N} \Big(
\op_{1a}^{N(2)} + \op_{2a}^{N(2)} \Big) + {\mathcal O}(q^5)\Big\}\,,
\end{split}
\\
\begin{split}\label{eq:Q9q7:had}
\Q_{9,q}^{(7)}\to&  8 F_{T,0}^{q/N}\op_4^N+\Big\{\Big[2 F_{T,1}^{q/N}\frac{\vec q^{\,\,2}}{m_N^2}
+F_{T,0}^{q/N}\Big(\frac{\vec q^{\,\,2}}{m_\chi^2} + \frac{\vec q^{\,\,2}}{m_N^2}-8\vasq\Big)\Big]\op_4^N+4 F_{T,0}^{q/N} \op_{2b}^N
\\
&-\frac{\vec q^{\,\,2}}{2 m_N m_\chi}\big(F_{T,0}^{q/N}-F_{T,1}^{q/N}\big)\op_1^N
-\Big[\Big(1+\frac{m_N^2}{m_\chi^2}\Big) F_{T,0}^{q/N}+2 F_{T,1}^{q/N}\Big]\op_6^N 
\\
&  -\frac{m_N}{m_\chi} F_{T,0}^{q/N}\op_3^N+ 2 \big(F_{T,0}^{q/N}-F_{T,1}^{q/N}\big) \op_5^N+ 16 F_{T,0}^{q/N}
\op_{3a}^{N(2)} + {\mathcal O}(q^3)\Big\}\,,
\end{split}
\\
\begin{split}\label{eq:Q10q7:had}
\Q_{10,q}^{(7)}\to&  -2 \frac{m_N}{m_\chi} F_{T,0}^{q/N}\op_{10}^N
+ 2 \big(F_{T,0}^{q/N}-F_{T,1}^{q/N}\big)\op_{11}^N
-8 F_{T,0}^{q/N}\op_{12}^N
\\
&  -\Big\{ \frac{m_N}{m_\chi} F_{T,0}^{q/N}\op_{10}^N\Big(\frac{\vec
  q^{\,\,2}}{4 m_\chi^2}+\vec v_\perp^{\,\,2}\Big)+8
F_{T,0}^{q/N}\op_{12}^N\Big(\frac{1}{2}\vec v_a^{\,\,2}
+\frac{1}{2}\vec v_\perp^{\,\,2} +\frac{\vec
  q^{\,\,2}}{8 m_\chi^2}\Big) 
\\
&+\op_{11}^N \Big[\Big(\frac{\vec q^{\,\,2}}{4 m_\chi^2} + \vec
  v_\perp^{\,\,2}\Big)F_{T,1}^{q/N} - F_{T,0}^{q/N} \Big(3\vec
  v_a^{\,\,2} + \vec v_\perp^{\,\,2} + \frac{\vec q^{\,\,2}}{4
    m_\chi^2}  + \frac{\vec q^{\,\,2}}{4
    m_N^2} \Big)
\\
 & + F_{T,2}^{q/N} \Big( 4 \vec
   v_a^{\,\,2} - \frac{\vec q^{\,\,2}}{m_N^2} \Big)\Big]
- \big(F_{T,0}^{q/N}+2F_{T,1}^{q/N}\big)\op_{15}^N
- 2 \frac{i \vec q \ncdot \vec v_a}{m_\chi} F_{T,0}^{q/N}
\op_{1a}^{N(1)} 
\\
&- \frac{i \vec q \ncdot \vec v_a}{m_N} \big(
2 F_{T,0}^{q/N} - F_{T,1}^{q/N}\big)\op_{2a}^{N(1)} 
+ 4 F_{T,0}^{q/N} \big( \op_{1a}^{N(3)} + \op_{2a}^{N(3)} \big) + {\mathcal O}(q^4)\Big\}\,.
\end{split}
\end{align}
The expressions that appear for the first time at ${\mathcal
  O}(q^{\nu_{\rm LO}+2})$ are collected inside the curly brackets. In
these the form factors are to be expanded to LO in chiral counting, as
denoted in Eqs. \eqref{eq:F_PP'}-\eqref{eq:Fi}. In particular, the
form factors without light meson poles are evaluated at $q^2=0$, i.e.,
for these $F_i\to F_i(0)$ inside curly brackets. In the terms outside
curly brackets, however, the form factors should be expanded to NLO,
cf. Eqs. \eqref{eq:F_PP'}-\eqref{eq:Fi}. The operators
$\Q_{5,q}^{(7)}$ and $\Q_{6,q}^{(7)}$ receive contributions at
${\mathcal O}(q^{\nu_{\rm LO}+1})$ from two-body currents, so we do
not display the corrections from expanding the single-nucleon currents
which, in this case, start at ${\mathcal O}(q^{\nu_{\rm LO}+2})$.

\section{Nonrelativistic expansion for scalar DM}
\label{app:NR:scalar}
To derive the HDMET for scalar DM, we factor out\footnote{Note that we
  dropped a global rescaling factor $(2m_\varphi)^{-1/2}$ on the right
  side of Eq.~\eqref{eq:redef:scal}.} the large momenta, 
\begin{equation}\label{eq:redef:scal}
\varphi(x)=e^{-i m_\varphi v\cdot x}\varphi_v\,,
\end{equation}
followed by a field redefinition 
\begin{equation}
\varphi_v\to \Big(1-i\frac{v\ncdot \partial}{4m_\varphi}+\frac{(i \partial_\perp)^2}{8 m_\varphi^2}+\frac{3}{32} \frac{(i v\ncdot \partial)^2}{ m_\varphi^2}-\frac{3}{32}\frac{(iv\ncdot \partial)(i\partial_\perp)^2}{ m_\varphi^3}-\frac{5}{128}\frac{(i v\ncdot \partial)^3}{m_\varphi^3}+\cdots \Big)\varphi_v\,.
\end{equation}
This gives the usual  HDMET for scalar DM 
\begin{equation}
{\cal L}_{\rm HDMET}=\varphi_v^* i v\ncdot \partial
\varphi_v^{\phantom{*}}+\frac{1}{2 m_\varphi}\varphi_v^*
(i\partial_\perp)^2\varphi_v^{\phantom{*}}+\frac{1}{8 m_\varphi^3}\varphi_v^*
(i\partial_\perp)^4\varphi_v^{\phantom{*}}+\cdots+{\cal
  L}_{\varphi_v}\,. 
\end{equation}
The first term is the LO HDMET for scalar fields. The $1/m_{\varphi}$
term is fixed by reparametrization invariance \cite{Luke:1992cs},
while the ellipses denote the higher-order terms. 

The DM bilinears have the following nonrelativistic expansion,
\begin{align}
\varphi^* \varphi & \to \varphi_v^* \varphi_v-\frac{1}{4m_\varphi^2}\varphi_v^* \big(\lpartial^2+\rpartial^2)\varphi_v+{\mathcal O}(q^3)\,,\label{eq:phiphi-NR}
\\
i\big(\varphi^* \overset{\leftrightarrow}{\partial_\mu} \varphi\big) & \to 2 m_\varphi v_\mu 
\big(\varphi_v^*  \varphi_v\big) +i\big(\varphi_v^* \overset{\leftrightarrow}{\partial}_{\perp,\mu} \varphi_v\big)+{\mathcal O}(q^3)\,, \label{eq:phidphi-NR}
\\
\big(\partial^{[\mu}\varphi^*\partial^{\nu]} \varphi\big) & \to i m_\varphi v^{[\mu}\partial_{\perp}^{\nu]}
\big(\varphi_v^*  \varphi_v\big) +\partial_{\perp}^{[\mu}\varphi_v^* \partial_{\perp}^{\nu]} \varphi_v+\frac{i}{4 m_\varphi} 
\varphi_v^* v^{[\mu}\lrpartial_{\perp}^{\nu]}\big(\lpartial^2-\rpartial^2) \varphi_v+{\mathcal O}(q^4)\,. \label{eq:dphidphi-NR}
\end{align}
In terms of the momenta these are
\begin{align}
\varphi^* \varphi & \to \varphi_v^* \varphi_v\Big(1+\frac{p_1^2+p_2^2}{4m_\varphi^2}\Big)+\cdots\,,\label{eq:phiphi-NR:mom}
\\
i\big(\varphi^* \overset{\leftrightarrow}{\partial_\mu} \varphi\big) & \to  
\varphi_v^*  \varphi_v \Big(2 m_\varphi v_\mu+p_{12\perp,\mu}\Big) +\cdots\,, \label{eq:phidphi-NR:mom}
\\
\big(\partial^{[\mu}\varphi^*\partial^{\nu]} \varphi\big) & \to
m_\varphi \bigg( v^{[\mu} q^{\nu]}
 + v^{[\mu} p_{12}^{\nu]} \frac{p_1^2-p_2^2}{4 m_\varphi^2} \bigg) \varphi_v^*  \varphi_v
+p_{2}^{[\mu} p_{1}^{\nu]} \varphi_v^* \varphi_v+\cdots\,. \label{eq:dphidphi-NR:mom}
\end{align}
The nonrelativistic reductions of the operators describing
interactions with scalar DM are thus (again explicitly including a
normalization factor similar to~\eqref{eq:norm})
\begin{align}
\begin{split}\label{eq:Q1:scal}
\Q_{1q}^{(6)}\to& 2 m_\varphi F_1^{q/N}\op_1^N\Big(1+\frac{\vec
  v_\perp^{\,\,2}}{2} + \frac{\vec q^{\,\,2}}{8 m_\varphi^2} \Big)-
 \frac{\vec q^{\,\,2}}{2 m_N^2} m_\varphi F_2^{q/N} \op_1^N
\\
&+2 m_\varphi \big(F_1^{q/N}+F_2^{q/N}\big)\op_3^N+{\mathcal O}(q^3)\,,
\end{split}
\\
\begin{split}\label{eq:Q2:scal}
\Q_{2q}^{(6)}\to &- 4 F_A^{q/N}m_\varphi \op_7^N\Big[1+\frac{\vec
  v_a^{\,\,2}}{2}+\frac{\vec
  v_\perp^{\,\,2}}{2} + \frac{\vec
    q^{\,\,2}}{8}\Big(\frac{1}{m_N^2}+\frac{1}{m_\varphi^2}\Big)\Big] 
\\
&- 2 F_A^{q/N}
m_\varphi (\vec v_a \ncdot \vec v_\perp) \op_{1a}^{N(1)} + {\mathcal
  O}(q^4)\,,
\end{split}
\\
\Q_{3q}^{(6)}\to & F_S^{q/N} \op_1^N\Big(1 + \frac{\vec q^{\,\,2}}{8 m_N^2}\Big)+{\mathcal O}(q^4)\,,\label{eq:Q3:scal}
\\
\Q_{4q}^{(6)}\to & F_P^{q/N} \op_{10}^N + \frac{1}{2}F_P^{q/N}\frac{(i
  \vec q\ncdot \vec v_a)}{m_N}\op_{1a}^{N(1)}+{\mathcal O}(q^4)\,,\label{eq:Q4:scal}
\\
\Q_{5}^{(6)}\to & F_G \op_1^N\Big(1 + \frac{\vec q^{\,\,2}}{8 m_N^2}\Big)+{\mathcal O}(q^4)\,,\label{eq:Q5:scal}
\\
\Q_{6}^{(6)}\to & F_{\tilde G} \op_{10}^N+\frac{1}{2}F_{\tilde G}\frac{(i \vec q\ncdot \vec v_a)}{m_N}\op_{1a}^{N(1)}+{\mathcal O}(q^4)\,,\label{eq:Q6:scal}
\end{align}
where the non-relativistic operators are defined in Eqs. \eqref{eq:O1pO2p}-\eqref{eq:O15p} and Eqs. \eqref{eq:O1aO1a1}-\eqref{eq:O3a3}. 

\section{The expressions for the non-relativistic coefficients}\label{sec:anand}

Here we collect the expressions for the coefficients of the
non-relativistic operators, Eqs.~\eqref{eq:O1pO2p}-\eqref{eq:O15p}, in
terms of the UV Wilson coefficients, Eq.~\eqref{eq:lightDM:Lnf5}, and
the single-nucleon form factors. We find
\begin{align}
\label{eq:cNR1}
c_{1}^p &=-\frac{\alpha}{2\pi m_\chi} Q_p \hat \C_1^{(5)}+ \sum_q \Big( F_{1}^{q/p} \, \hat
\C_{1,q}^{(6)} + F_{S}^{q/p} \, \hat \C_{5,q}^{(7)} \Big) + F_G^p \,
\hat \C_{1}^{(7)} 
\\
&\qquad - \frac{\vec q^{\,\,2}}{2m_\chi m_N} \sum_q \big( 
F_{T,0}^{q/p} - F_{T,1}^{q/p} \big) \hat \C_{9,q}^{(7)} \,,
\\
\label{eq:cNR4}
c_{4}^p &= -\frac{2\alpha}{\pi}\frac{\mu_p}{m_N} \hat\C_1^{(5)}+ \sum_q \Big( 8 F_{T,0}^{q/p} \, \hat
\C_{9,q}^{(7)} - 4 F_{A}^{q/p} \, \hat \C_{4,q}^{(6)} \Big) \,,
\\
\label{eq:cNR5}
c_{5}^p &=\frac{2\alpha Q_p m_N}{\pi \vec q^{\,\,2}}\hat \C_1^{(5)} +
2 \big( F_{T,0}^{q/p} - F_{T,1}^{q/p} \big) \hat \C_{9,q}^{(7)} \,,
\\
\label{eq:cNR6}
c_{6}^p &= \frac{2\alpha}{\pi\vec q^{\,\,2}}\mu_p m_N\hat \C_1^{(5)}+\sum_q \Big( F_{P'}^{q/p} \, \hat
\C_{4,q}^{(6)} + \frac{m_N}{m_\chi} F_{P}^{q/p} \, \hat \C_{8,q}^{(7)}
\Big) + \frac{m_N}{m_\chi} F_{\tilde G}^{p} \, \hat \C_{4}^{(7)} \,,
\\
\label{eq:cNR7}
c_{7}^p &= - 2 \sum_q F_{A}^{q/p} \, \hat \C_{3,q}^{(6)} \,,
\\
\label{eq:cNR8}
c_{8}^p &= 2 \sum_q F_{1}^{q/p} \, \hat \C_{2,q}^{(6)} \,,
\\
\label{eq:cNR9}
c_{9}^p &= 2 \sum_q \Big[ \big( F_{1}^{q/p} + F_{2}^{q/p}
\big) \, \hat \C_{2,q}^{(6)} + \frac{m_N}{m_\chi} F_{A}^{q/p} \, \hat
\C_{3,q}^{(7)} \Big] \,,
\\
\label{eq:cNR10}
c_{10}^p &= F_{\tilde G}^p \, \hat \C_{3}^{(7)} + \sum_q
\Big( F_{P}^{q/p} \, \hat \C_{7,q}^{(7)} - 2 \frac{m_N}{m_\chi}
F_{T,0}^{q/p} \, \hat \C_{10,q}^{(7)} \Big) \,,
\\
\label{eq:cNR11}
c_{11}^p &= \frac{2\alpha}{\pi}Q_p \frac{m_N}{\vec q^{\,\,2}}\hat \C_2^{(5)}+ \sum_q \Big[ 2 \big( F_{T,0}^{q/p} -
  F_{T,1}^{q/p} \big) \, \hat \C_{10,q}^{(7)} - \frac{m_N}{m_\chi}
  F_{S}^{q/p} \, \hat \C_{6,q}^{(7)} \Big] - \frac{m_N}{m_\chi}
F_{G}^p \, \hat \C_{2}^{(7)} \,, 
\\ 
c_{12}^p&= - 8 \sum_q
F_{T,0}^{q/p} \, \hat \C_{10,q}^{(7)} \,.
\end{align}
The coefficients for neutrons are obtained by replacing $p\to n$,
$u\leftrightarrow d$.  Above we kept only the chirally leading
contributions and listed the results only for the non-vanishing
$c_{{\rm NR},i}^N$ (i.e., one has $c_{{\rm NR},2}^N=c_{{\rm
    NR},3}^N=0$). For the coefficient $c_{{\rm NR},1}^N$, we also kept
the $q^2$-suppressed contribution from $\hat \C_{9,q},^{(7)}$ that is,
however, coherently enhanced. The contributions due to the magnetic
and electric dipole operators, Eqs.~\eqref{eq:dim5:nf5:Q1Q2:light},
are given in Appendix~A of~\cite{Bishara:2016hek}.

In the LO expressions most of the form factors are evaluated at
$q^2=0$, with the numerical values for $F_{1}^{q/N}$ given in
Eq.~\eqref{eq:F1:num}; for $F_{2}^{q/N}$ in
Eqs.~\eqref{eq:F2s:num}-\eqref{eq:F2d:num}; for $F_{A}^{q/N}$ in
Eq.~\eqref{eq:FA:def} together with Eqs.~\eqref{eq:Deltau:def},
\eqref{eq:Deltaq:values}; for $F_{S}^{q/N}$ in Eq.~\eqref{eq:FS:def}
together with Eqs.~\eqref{eq:values:sigmas},
\eqref{eq:values:sigmaqN}; for $F_{G}^{q/N}$ in Eq.~\eqref{eq:FG:def}
together with~\eqref{eq:mG:value}; for $F_{T,0}^{q/N}$ in
Eq.~\eqref{eq:FT0:tqN} together with~\eqref{eq:tqN:val}; and for
$F_{T,1}^{q/N}$ in Eq.~\eqref{eq:FT1:def} together
with~\eqref{eq:BT10u}-\eqref{eq:BtildeA:s}.  The form factors
$F_{P}^{q/N}$, $F_{P'}^{q/N}$, $F_{\tilde G}^{q/N}$ contain pion and
eta poles. The numerical values for $F_{P'}^{q/N}$ are given in
Eq.~\eqref{eq:F_P':LO} together with Eqs.~\eqref{eq:aP'pi},
\eqref{eq:aP'eta},~\eqref{eq:App:u-d},~\eqref{eq:Deltaq:values}; for
$F_{P}^{q/N}$ in Eq.~\eqref{eq:F_P:LO}-\eqref{eq:B0mq:num} together
with Eqs.~\eqref{eq:App:u-d},~\eqref{eq:Deltaq:values}; for $F_{\tilde
  G}^{q/N}$ in Eq.~\eqref{FGtilde:LO} together with
~\eqref{eq:App:u-d},~\eqref{eq:Deltaq:values}.

\bibliography{NLO_paper}

\end{document}